\documentclass[final,3p,times,twocolumn]{elsarticle}
\usepackage{hyperref}
\usepackage{amssymb}
\usepackage{lineno}
\usepackage{caption}
\usepackage{subcaption}
\usepackage{siunitx}
\journal{Nuclear Instruments and Methods in Physics A}

\graphicspath{{Figures/}}

\begin{document}
\begin{frontmatter}
\title{The Mini-CAPTAIN Liquid Argon Time Projection Chamber}


\author[label1]{C. E. Taylor\corref{mycorrespondingauthor}}
\cortext[mycorrespondingauthor]{Corresponding author. \emph{Email:} cetaylor@lanl.gov.}
\author[label4]{B. Bhandari}
\author[label8]{J. Bian}
\author[label5]{K. Bilton}
\author[label3]{C. Callahan}
\author[label3]{J. Chaves}
\author[label13]{H. Chen}
\author[label14]{D. Cline\fnref{Deceased}}
\fntext[Deceased]{Deceased}
\author[label2]{R. L. Cooper}
\author[label1,label5]{D.~L.~Danielson\fnref{myfootnote}}
\fntext[myfootnote]{Present address:  Enrico Fermi Institute \& Department of Physics, The University of Chicago, Chicago, Illinois 60637, USA}
\author[label1]{J. Danielson}
\author[label6]{N. Dokania}
\author[label1]{S. Elliot}
\author[label15]{S. Fernandes}
\author[label5]{S. Gardiner\fnref{myfootnote2}}
\fntext[myfootnote2]{Present address:  Fermi National Accelerator Laboratory, Batavia, Illinois 60510, USA}
\author[label1]{G. Garvey}
\author[label9]{V. Gehman}
\author[label7]{F. Giuliani}
\author[label3]{S. Glavin}
\author[label7]{M. Gold}
\author[label11]{C. Grant}
\author[label1]{E. Guardincerri}
\author[label1]{T. Haines}
\author[label4]{A. Higuera\fnref{myfootnote3}}
\fntext[myfootnote3]{Present address:  Department of Physics and Astronomy, Rice University, Houston, TX 77005, USA}
\author[label6]{J. Y. Ji}
\author[label9]{R. Kadel}
\author[label8a]{N. Kamp}
\author[label3]{A. Karlin}
\author[label1]{W. Ketchum}
\author[label4]{L. W. Koerner}
\author[label1]{D. Lee}
\author[label14]{K. Lee}
\author[label1]{Q. Liu}
\author[label8]{S. Locke}
\author[label1]{W. C. Louis}
\author[label9]{P. Madigan}
\author[label5]{A. Manalaysay}
\author[label10]{J. Maricic}
\author[label14]{E. Martin}
\author[label1]{M. J. Martinez}
\author[label6]{S. Martynenko}
\author[label3]{C. Mauger}
\author[label6]{C. McGrew}
\author[label1]{J. Medina}
\author[label1]{P. J. Medina\fnref{Deceased}}
\author[label1]{G. Mills\fnref{Deceased}}
\author[label1]{J. Mirabal-Martinez}
\author[label17]{A. Olivier}
\author[label5]{E. Pantic}
\author[label7]{B. Philipbar}
\author[label8]{C. Pitcher}
\author[label13]{V. Radeka}
\author[label1]{J. Ramsey}
\author[label1]{K. Rielage}
\author[lable10]{M. Rosen}
\author[label1]{A. R. Sanchez}
\author[label14]{J. Shin}
\author[label1]{G. Sinnis}
\author[label8]{M. Smy}
\author[label1]{W. Sondheim}
\author[label15]{I. Stancu}
\author[label1]{C. Sterbenz}
\author[label10]{Y. Sun}
\author[label5]{R. Svoboda}
\author[label14]{A. Teymourian}
\author[label13]{C. Thorn}
\author[label9]{C. E. Tull}
\author[label17]{M. Tzanov}
\author[label1]{R. Van de Water}
\author[label5]{N. Walsh}
\author[label14]{H. Wang}
\author[label14]{Y. Wang}
\author[label6]{C. Yanagisawa}
\author[label1]{A. Yarritu}
\author[label4]{J. Yoo}
  
 \address[label1]{Los Alamos National Laboratory, Los Alamos, New Mexico 87545, USA}
 \address[label2]{Department of Physics, New Mexico State University, Las Cruces, NM 88003, USA}
 \address[label3]{Department of Physics and Astronomy, University of Pennsylvania, Philadelphia, PA 19104, USA}
 \address[label4]{Department of Physics, University of Houston, Houston, Texas 77204, USA} 
 \address[label5]{Department of Physics and Astronomy, University of California, Davis, CA 95616, USA}
 \address[label6]{Department of Physics and Astronomy, Stony Brook University, Stony Brook, NY 11794, USA}
 \address[label7]{Department of Physics and Astronomy, University of New Mexico, Albuquerque, NM 87131, USA}
 \address[label8]{Department of Physics and Astronomy, University of California, Irvine, CA 92697, USA}
 \address[label8a]{Department of Physics, University of Michigan, Ann Arbor, Michigan 48109, USA}
 \address[label9]{Lawrence Berkeley National Laboratory, Berkeley, CA 94720, USA}
 \address[label10]{Department of Physics and Astronomy, University of Hawaii at Manoa, Honolulu, HI 96822, USA}
 \address[label11]{Department of Physics, Boston University, Boston, MA 02215, USA}
 \address[label13]{Brookhaven National Laboratory, Upton, NY 11973, USA}
 \address[label14]{Department of Physics and Astronomy, University of California, Los Angeles, CA 90095, USA}
 \address[label15]{Department of Physics and Astronomy, University of Alabama, Tuscaloosa, AL 35487, USA}
 \address[label17]{Department of Physics and Astronomy, Louisiana State University, Baton Rouge, LA 70803, USA}
 \address[label18]{Fermi National Accelerator Laboratory, Batavia, Illinois 60510, USA}
 
\begin{abstract}
This manuscript describes the commissioning of the Mini-CAPTAIN liquid argon detector in a neutron beam at the Los Alamos Neutron Science Center (LANSCE), which led to a first measurement of high-energy neutron interactions in argon.  The Mini-CAPTAIN detector consists of a Time Projection Chamber (TPC) with an accompanying photomultiplier tube (PMT) array sealed inside a liquid-argon-filled cryostat.  The liquid argon is constantly purified and recirculated in a closed-loop cycle during operation. The specifications and assembly of the detector subsystems and an overview of their performance in a neutron beam are reported.


\end{abstract}

\end{frontmatter}
\section{Introduction} \label{sec:introduction}
A key aspect of long-baseline neutrino oscillation measurements is the ability to infer the neutrino flavor ($e$, $\mu$, or $\tau$) and type ($\nu$ or $\overline{\nu}$).  This requires clear identification of the outgoing lepton coming from the neutrino interaction, such as $e^-$ or $e^+$ in the case of a $\nu_e$ or $\overline{\nu}_e$. With this in mind, liquid argon time-projection chambers (LArTPCs) provide several advantages over other detectors, like those based on water Cherenkov technology. A combination of $dE/dx$ differentiation and high-resolution track topology of final state particles in LArTPCs allow for improved particle identification, thereby suppressing backgrounds.  
ICARUS was one of the first experiments to successfully demonstrate the advantages of using a LArTPC~\cite{bib:ICURUS}. 

The Cryogenic Apparatus for Precision Tests of Argon Interactions with Neutrinos (CAPTAIN) project was conceived to investigate the technical challenges of developing large-volume LArTPCs and explore their physics capabilities~\cite{bib:CaptProposal}.  Mini-CAPTAIN is a first prototype of CAPTAIN and it made a first measurement of the high-energy neutron cross-sections and event signatures in a LArTPC~\cite{bib:neutronMeas}.  In addition to a cross-section measurement, Mini-CAPTAIN also tested a number of hardware advancements for future LArTPCs.  The hardware advancements consisted of new TPC cold electronics developed by Brookhaven National Lab (BNL) for the MicroBooNE detector, a prototype hexagonal time-projection chamber (TPC) designed for the larger CAPTAIN detector, a chemically-etched cathode grid plane, and several configurations for injecting high-power laser light into the TPC fiducial volume.  This report details the specifications and assembly of the detector subsystems and provides an overview of their performance in a neutron beam.
\section{Mini-CAPTAIN Detector}\label{sec:hardware}

Mini-CAPTAIN consists of a hexagonal TPC with a 50\,cm apothem and a 32\,cm drift distance corresponding to a total active liquid argon mass of 400\,kg.  A high-voltage cathode wire plane is used to apply an electric field of 500\,V/cm across the active volume of the TPC which corresponds to an electron drift velocity of 1.6\,mm/$\mu$s.  Ionized electrons drift away from the cathode plane and towards a series of sensing wire planes where they form signals for readout. Accompanying the TPC is a photon detection system (PDS) consisting of an array of photomultiplier tubes (PMTs).  The PMTs view the active region of the TPC through wavelength shifting plates positioned below the cathode plane (looking up) and positioned above the wire sensing planes (looking down). The PDS is designed to collect the vacuum ultraviolet scintillation light in conjunction with the ionized charge measured by the TPC.  The addition of scintillation light aids in event reconstruction, provides timing for non-beam events related to cosmic rays or radioactivity from the surrounding environment, and provides a measurement of neutron kinetic energy via time-of-flight.

\begin{figure}[t!]
	\centering
        \includegraphics[width=6.0cm]{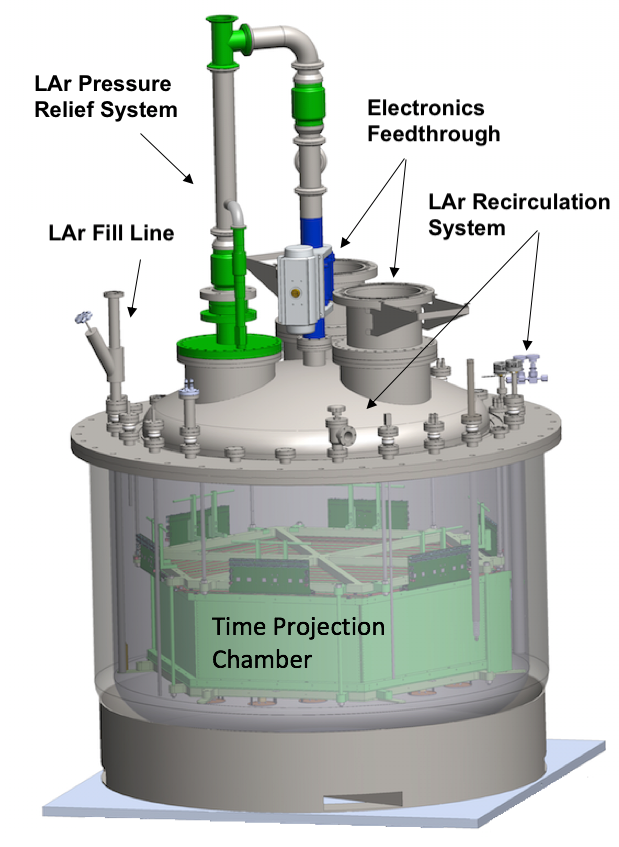}
    \caption{Overview of the Mini-CAPTAIN detector.  
    }
        \label{fig:miniCaptainDesign}
\end{figure}

The TPC and PDS are contained inside a large cryostat. All instrumentation and cryogenic leads are fed through the top lid of the cryostat.  The flanges on the outside of the top lid allow access to the PDS, liquid argon purity monitor, and re-circulation heater.  It also facilitates optical access to the liquid argon for laser calibration. The entire detector assembly was positioned on stationary blocks in the neutron beam line.  An overview drawing of the Mini-CAPTAIN detector is shown in Figure~\ref{fig:miniCaptainDesign}.  Extensive details and specifications of all of the Mini-CAPTAIN detector subsystems are described in the sections that follow.

\subsection{Cryogenic Subsystems} \label{sec:cyro_config}

Liquid argon has proven to be a cost-effective material to serve both as the target and detector medium\cite{bib:ICURUS, bib:Anderson} due to its relatively large natural abundance and the ease at which it can be refrigerated with inexpensive liquid nitrogen. As a detector medium, charged particles traversing a liquid argon volume will leave behind trails of ionized electrons and produce an abundance of scintillation light at vacuum ultraviolet wavelengths.  However, sufficient purity must be maintained in order to properly utilize these signals. Residual contamination in industrial liquid argon, such as nitrogen, water, and oxygen, can significantly impede the detector performance. For example, nitrogen degrades scintillation yield due to the addition of non-radiative processes that induce quenching.  Oxygen and water are electronegative impurities and will capture ionized electrons, thereby greatly reducing their lifetime and average drift distance. 

The TPC drift distance of 32\,cm requires that the concentrations of H$_2$O and O$_{2}$ be less than 0.3\,ppb and 1.5\,ppb, respectively.  The relative difference in H$_2$O and O$_{2}$ concentrations is due to the fact that with an electric field strength of 500\,V/cm, H$_{2}$O has an electron attachment rate five times stronger than that of O$_{2}$~\cite{bib:Bakale}.  Studies have shown that to achieve optimal scintillation yield, the N\textsubscript{2} contamination must be smaller than 2 ppm such that quenching and absorption of scintillation light are negligible~\cite{bib:Acciarri,bib:Jones}.  The maximum allowable H$_2$O and O$_2$ concentrations are roughly three orders of magnitude lower than the contamination level of the industrial liquid argon delivered to Los Alamos (H$_2$O and O$_2$ concentrations of 0.5 and 2.3\,ppm, respectively).  Therefore, extensive purification is required.  

The cryogenic systems comprise a cryostat, a gas recirculation, filtration and condensing system, and a criotec liquid purification system.  During filling, the argon was passed through an inline filter.  These subsystems are further discussed below.





\raggedbottom 


\subsubsection{Cryostat}
The cryostat is an evacuated dual-wall vessel made out of 304 stainless steel.  It has a height of 1.47\,m, an inner diameter of 1.40\,m and an outer diameter of 1.86\,m. The inner wall thickness is 0.16\,cm and the outer wall thickness is 0.32 cm.  The inner shell has a depth of 49\,cm. Though the TPC hangs 18.6 cm from the top of the cryostat, the liquid argon level was kept another 9\,cm higher to ensure that the cold electronics (see Section~\ref{sec:electronics_config}), sitting on top of the TPC, were submerged. 

\begin{figure}[b!]
	\centering
        \includegraphics[width=\linewidth]{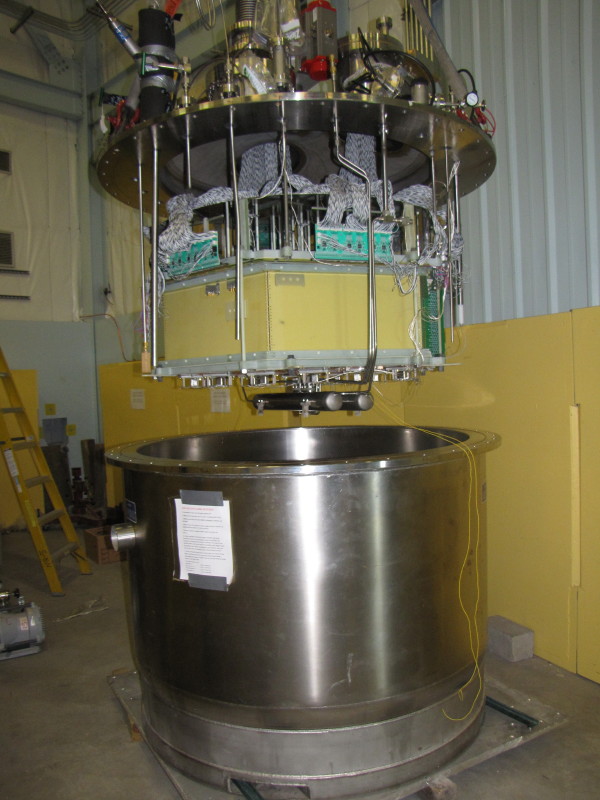}
    \caption{The Mini-CAPTAIN cryostat during its deployment into the WNR facility at the LANSCE accelerator facility. The TPC is suspended from the top head of the cryostat and the cold cables are shown connecting the motherboards and wire planes to the service boards on top of the lid. }
        \label{fig:miniCaptainCryo}
\end{figure}

The lid has multiple conflat flanges for the cryogenic systems, electronics feedthroughs, and laser calibration testing.  The cryogenic ports included a fill line, a gas recirculation port, liquid return ports, and a pressure relief valve. There are two large conflat flanges for the TPC feedthroughs and twenty-four 2.75 inch conflat flanges used for various systems such as the photon detection system and high voltage feedthrough.  Two additional ports for a laser calibration system were positioned between the two TPC feedthrough conflats. A heater was also installed for control of the boil-off rate and tank pressure.   Figure \ref{fig:miniCaptainCryo} shows Mini-CAPTAIN before its final deployment into the WNR facility of the LANSCE accelerator.

Two important factors impacting the purification of liquid argon are the cryostat leak rate and the outgassing of internal detector components. The purification system must remove impurities at a rate exceeding the introduction of contaminants from system leaks and component outgassing. Leaking can be reduced with proper sealing of the cryostat.  Proper selection of material used inside the detector can reduce the outgassing of oxygen and water. Diminished purity due to outgassing was further mitigated by implementing alternating cycles of vacuum pumping followed by argon gas purging before the cryostat is filled with liquid argon.  During this cycling period, the cryostat was evacuated to a level better than $10^{-5}$ Torr before being purged with pure argon gas.

\subsubsection{Liquid Argon Inline Filter System}
\begin{figure}[t!]
	\centering
        \includegraphics[width=0.65\linewidth]{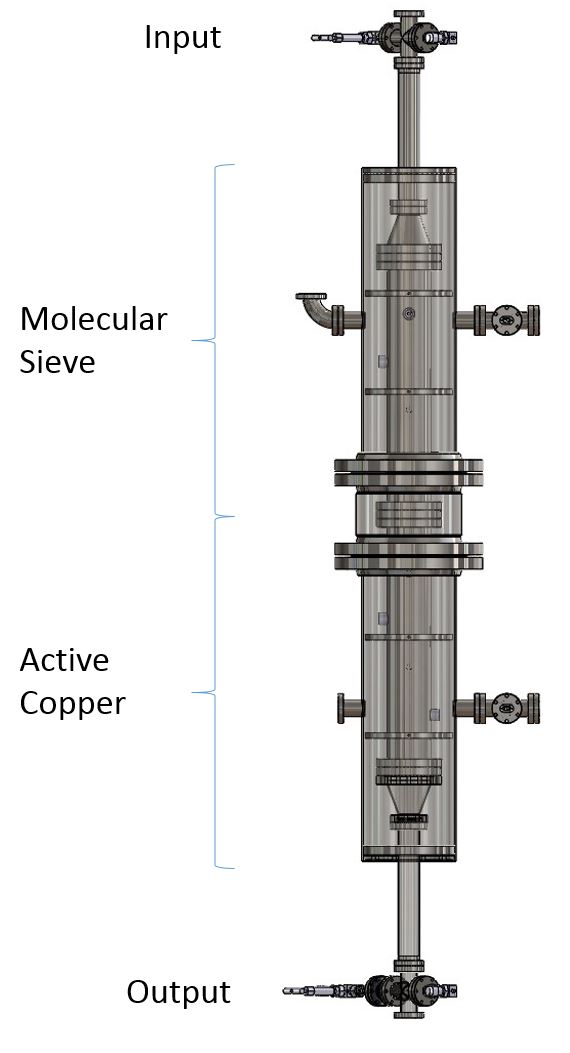}
    \caption{The Mini-CAPTAIN inline filter used for initial filling of liquid argon into the detector.}
        \label{fig:inlineFilter}
\end{figure}

During the initial filling of the cryostat, liquid argon is passed through an inline filter.  It has two active components: a molecular sieve to remove H$_2$O, and activated copper for removal of O$_2$. The inline filter is regenerated after each fill of the cryostat. As can be seen in Figure~\ref{fig:inlineFilter}, the filter is aligned in an upright position, such that the liquid argon is fed in through the top and extracted from the bottom. During regeneration, gaseous argon is circulated through the filter in the opposite direction.  The O$_2$ concentration is measured with a DF-550E Process Oxygen Analyzer from Servomex Company Inc., and the H$_2$O concentration is measured with a Continuous Wave CRDS Trace Gas Analyzer from Tiger Optics, LLC. These instruments sample the evaporated argon gas after the inline filter and have sensitivities for O$_2$ and H$_2$O of 3\,ppb and 1\,ppb, respectively.  The inline filter was observed to lower the O$_{2}$ concentration to 400\,ppb and the H$_{2}$O concentration to 500\,ppb.  A liquid recirculation system, described in the following section, was used further reduce the contaminates after filling. 

\subsubsection{Gaseous Argon Recirculation System}
Further purification of the argon in the cryostat is provided by a gas recirculation system.  The recirculation system collects gas from the cryostat ullage.  The gas is passed through two parallel SAES getter cartridges where oxygen and water are extracted.  Finally, the gas is then fed into the condenser system at the top of the recirculation tower before returning cryostat.

\begin{figure}[t!]
	\centering
        \includegraphics[width=\linewidth]{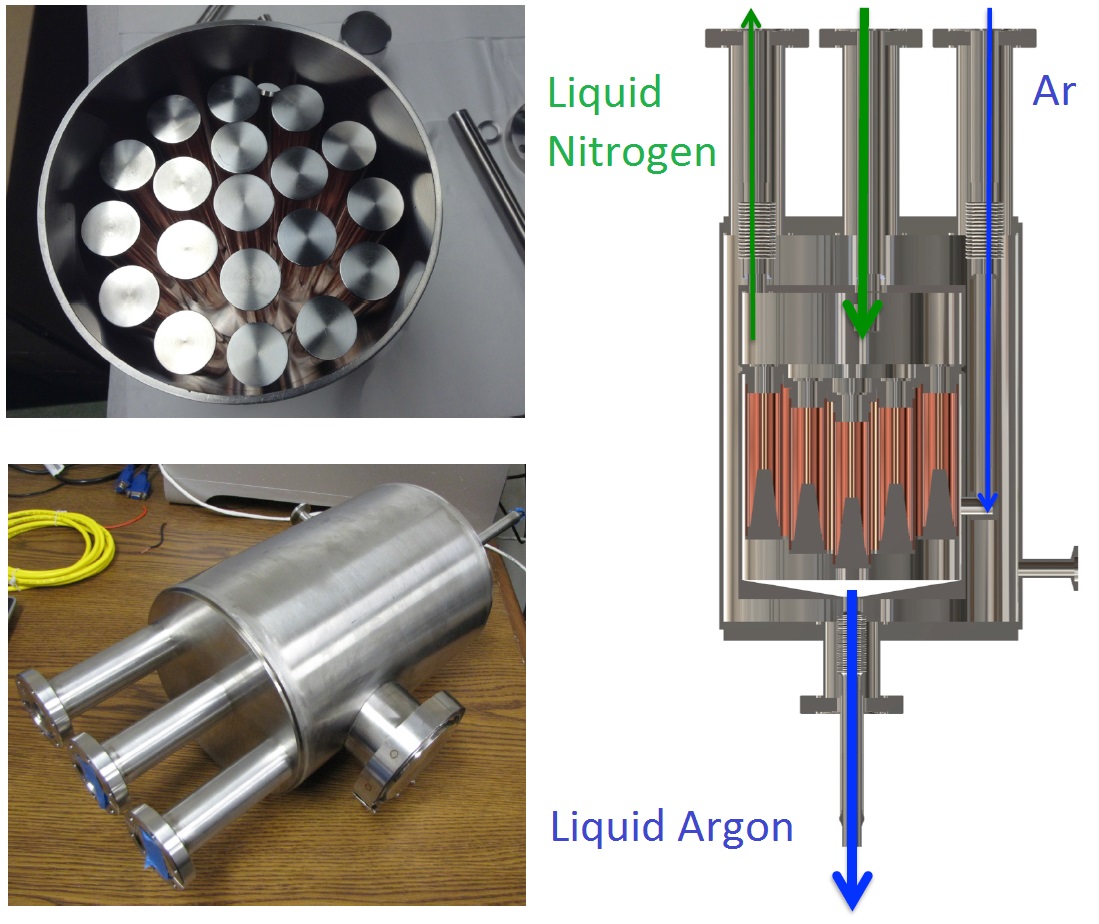}
    \caption{Two images of the liquid argon condenser system for Mini-CAPTAIN are shown on the left. A diagram of the general flow of argon through the condenser is shown on the right.}
        \label{fig:Condenser}
\end{figure}

The condenser was designed by Dr. Hanguo Wang at UCLA.  It consists of 19 copper tubes brazed with stainless steel, which have different lengths.  By varying the flow of N$_{2}$ the depth in each tube is controlled allowing the cold surface area in contact to vary. The evaporated argon gas flows into the top, where it runs between the tubes cooled by liquid nitrogen. After heat is transferred, the condensed argon drips to the bottom and falls back into the cryostat. Temperature sensors were used on the nitrogen input and output to observe the heat transfer during operation.  Two images showing the inside and outside of the condenser on the left side of Figure~\ref{fig:Condenser}.  A diagram depicting the flow of gaseous argon into the top of the condenser and flow of the liquid leaving the condenser is shown on the right side of Figure~\ref{fig:Condenser}.

During early stages of liquid argon filling, achieving the required liquid argon purity proved challenging, particularly due to lingering oxygen contamination. 
The previously described system ultimately achieved contamination levels of 3\,ppb for O$_{2}$ and 1.5\,ppb for H$_{2}$O.  The corresponding electron lifetime was deemed insufficient.  

\subsubsection{Criotec Filtration System}
\begin{figure}[b!]
	\centering
        \includegraphics[width=0.98\linewidth]{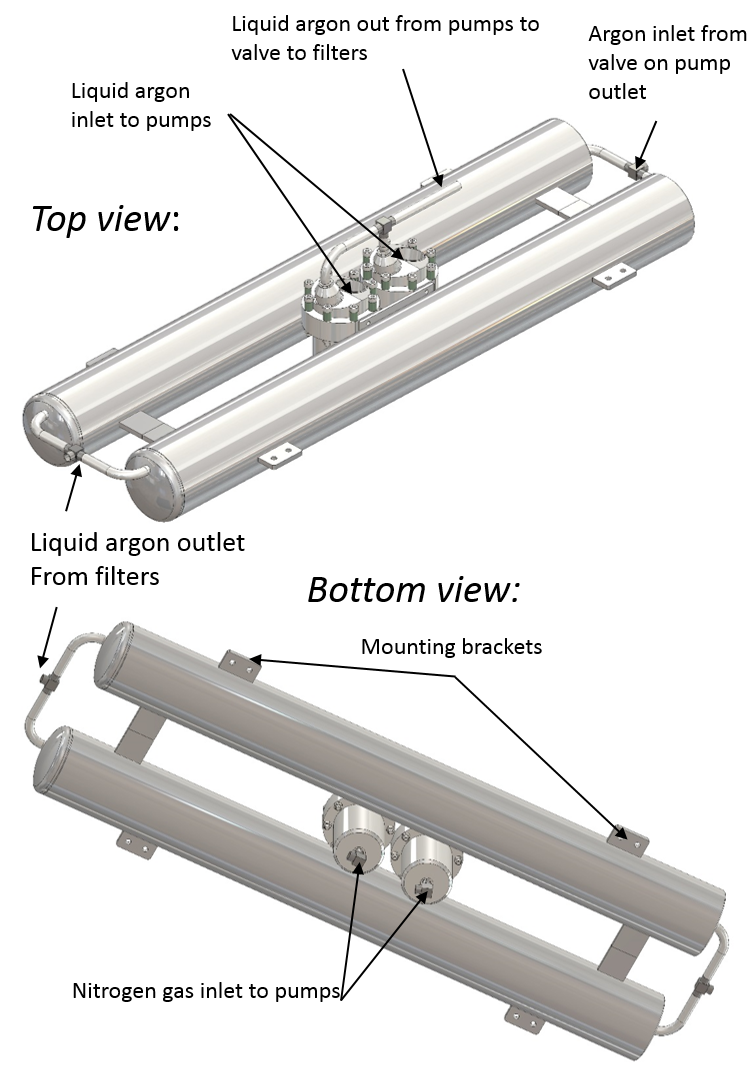}
    \caption{Diagram of the Criotec filtration system.}
        \label{fig:CriotecFilter}
\end{figure}

After tests demonstrated that the inline filter and recirculation system could not achieve the necessary purity, a new liquid argon purification system was procured from Criotec Impianti~\cite{bib:Criotec}. The system was designed with two purifier cylinders made with sintered copper-alumina, shown in Figure~\ref{fig:CriotecFilter}, and is attached directly underneath the TPC.  Liquid argon is cycled through the filter via the pressurization of two bellows pumps, which is achieved using a slight over-pressure of nitrogen gas. Two pneumatic timers alternate control over a 5/2 valve to feed the nitrogen over-pressure to the pumps. During the late stages of commissioning the Criotec filtration system brought the O$_2$ contamination level down to approximately 1\,ppb. 

\subsection{Time Projection Chamber} \label{sec:electronics_config}

The hexagonal LArTPC consists of four primary components: a cathode plane, a field cage, a stack of wire planes, and a ground plane. The cathode plane was mounted to the bottom of the field cage, while a series of five wire planes were mounted to the top. Ordered along the direction of electron drift, the planes are labeled as: grid, u, v, x  and ground. The u- and v-planes are rotated $\pm 60$ degrees, respectively, relative to the anode x-plane. Each wire plane consists of 332 wires made of 75\,$\mu$m diameter gold plated copper beryllium, with 3\,mm inter-wire spacing. The planes are separated by a vertical distance of 3.125\,mm. The wires are biased so that drifting electrons pass through the grid, u-, and v-planes unimpeded. The electrons are collected on the x plane. Induced signals are sensed on the u and v wires. The grid plane is used to define the induced signal on the u-plane to be determined by the space between between the grid and v-plane.



\begin{figure}[t!]
        \includegraphics[width=.48\textwidth]{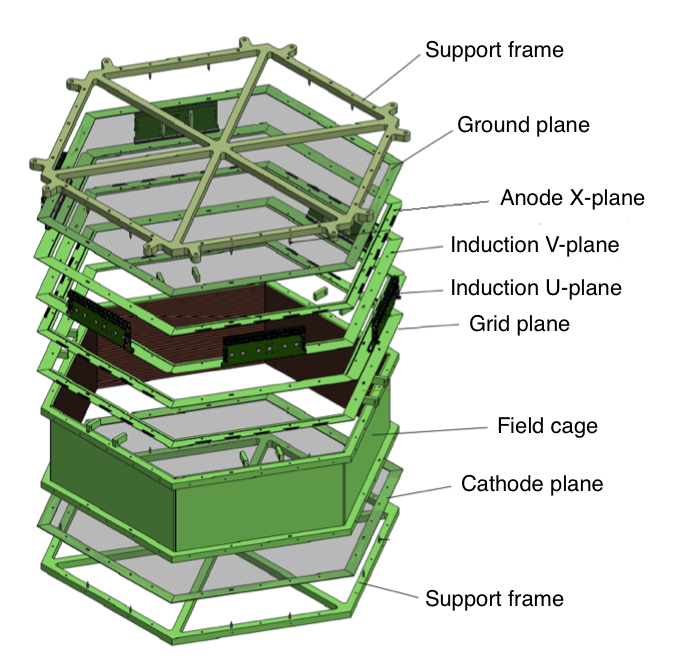}
        \caption{Diagram of the Mini-CAPTAIN TPC.}
        \label{fig:tpc_diagram}
\end{figure}

\begin{figure}[t!]
        \includegraphics[width=0.48\textwidth]{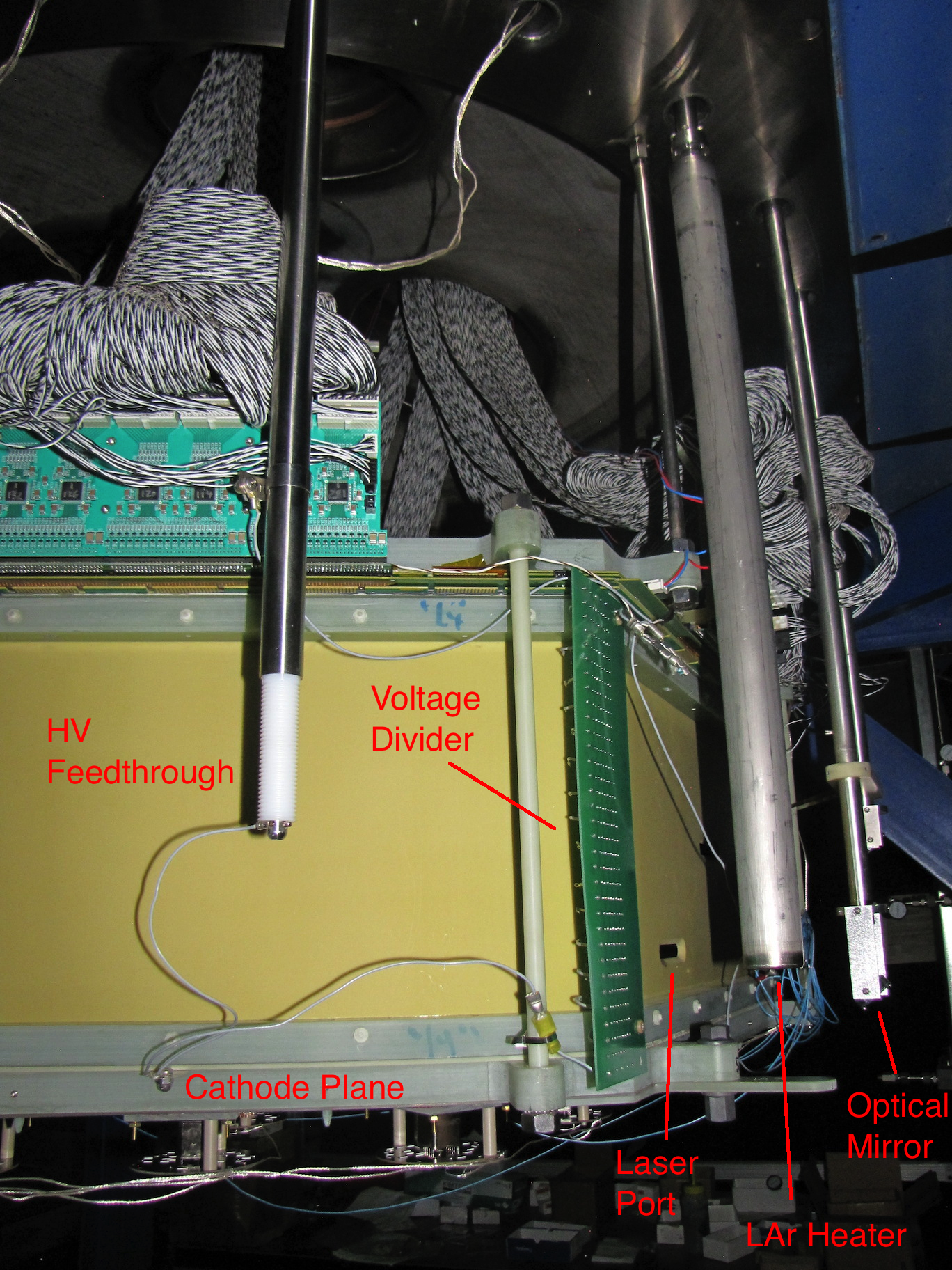}
        \caption{Close-up image showing the field cage assembly with the HV and calibration components labeled.}
        \label{fig:heater_cathode}
\end{figure}


The hexagonal field cage is constructed out of six copper clad FR4 sheets with a pattern on the inside of the cage that consists of 5-mm-wide strips separated by 10\,mm running from the cathode end to the wire plane end.  A resistive divider chain supplies high-voltage (HV) for each trace and gold-plated jumpers bridge the traces across each of the six cage corners. The resistive chain uses 25\,M$\Omega$/cm divisions, producing a gradient of 500\,V/cm for 20\,$\mu$A of current.  The chamber's bolts and nuts are made from G-10 fiberglass laminate.  Thick (2.54\,cm) FR4 support frames held tension on the wire planes during assembly, and also bound the LArTPC components together as the assembly hung from the top head of the cryostat.  

Two 1\,cm $\times$ 5\,cm slits were milled through one face of the field cage, providing a window into the active volume for a light beam from a Quantel Brilliant B Q-switched Nd:YAG laser calibration source.  The laser has a wavelength of 1064\,nm and pulse energy of 850\,mJ.  Frequency doublers are used to shift the wavelength to 266\,nm with a pulse energy of 90\,mJ.  The light is directed into the LArTPC using a periscope with an optical mirror in such a way as to span the LArTPC drift region, thereby creating well-defined ionization tracks within the LArTPC.  During the commissioning process the laser produced ionization tracks which were measured by the LArTPC wire planes thereby providing additional confirmation that the required liquid argon purity had been achieved.

A diagram of the LArTPC wire plan and field cage assembly is shown in Figure~\ref{fig:tpc_diagram}.  An image showing the relative positions and assembly of the HV feedthrough, voltage divider, liquid argon heater, and laser calibration port is shown in Figure~\ref{fig:heater_cathode}. 

\begin{figure*}[t!]
    \centering
    \begin{subfigure}{0.49\textwidth}
        \centering
        \includegraphics[height=2.2in]{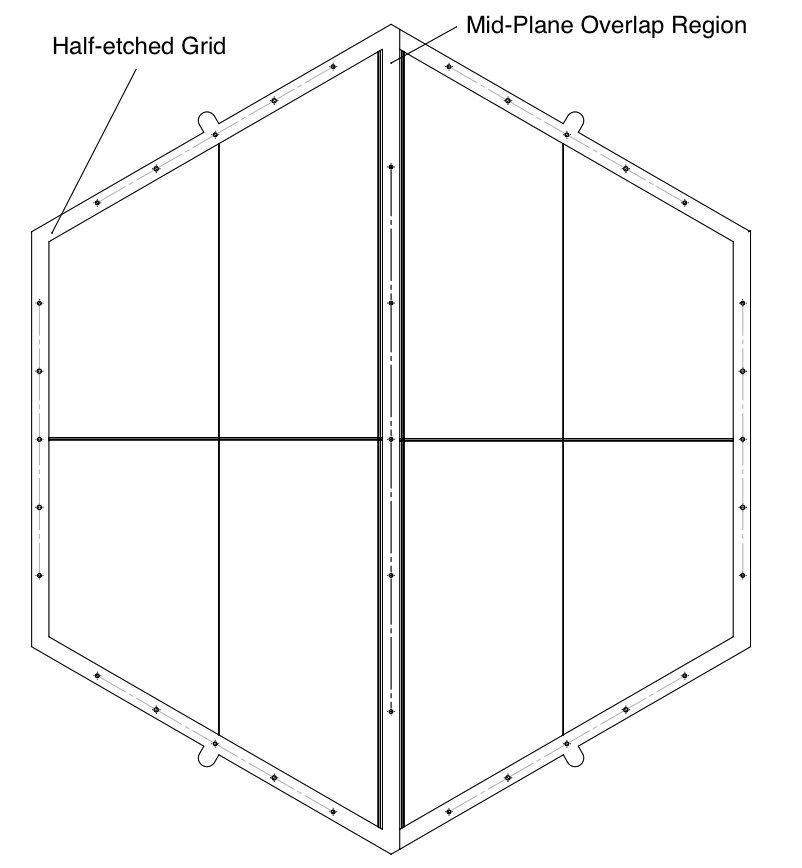}
        \caption{Diagram of Assembled Cathode}
        \label{fig:cathode_plane_a} 
    \end{subfigure}%
    \begin{subfigure}{0.49\textwidth}
        \centering
        \includegraphics[height=2.2in]{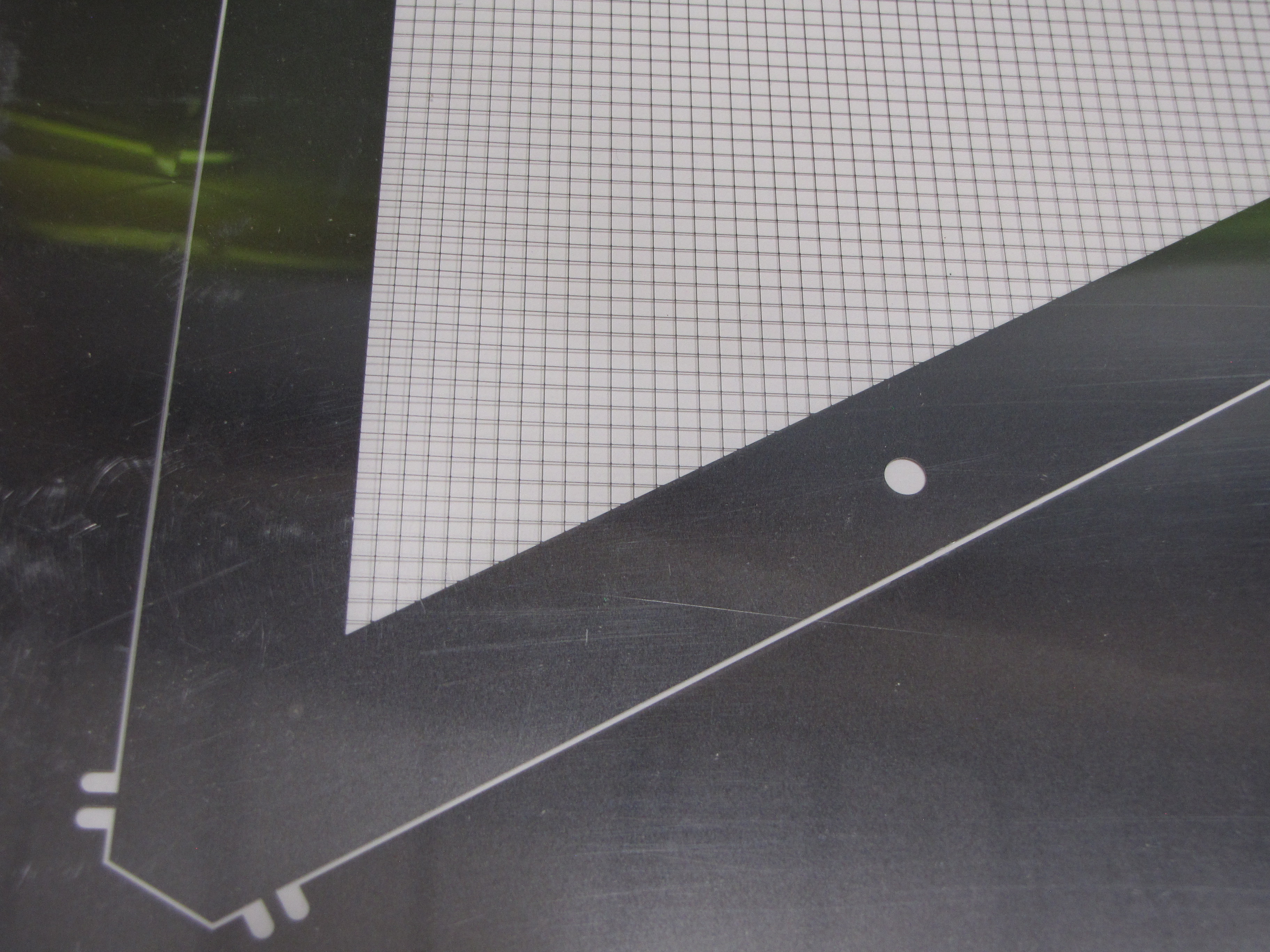}
        \caption{Close-Up of Etched Cathode}
        \label{fig:cathode_plane_b}
    \end{subfigure}

    \caption{A diagram of the assembled cathode plane is shown in (\subref{fig:cathode_plane_a}) and a close-up image of the etched grid on stainless steel sheet in shown in (\subref{fig:cathode_plane_b}).}
        \label{fig:cathode_plane}
\end{figure*}

\subsubsection{Cathode Plane}

A copper clad FR4 sheet with PMT windows acts as the ground plane and is described in more detail in Section~\ref{sec:pds_config}.  The cathode plane is formed from two chemically etched trapezoidal sheets of stainless steel with 2.56\,cm of overlap between them.  To the best of our knowledge, this is the first successful demonstration of a chemically etched cathode plane instrumented inside a TPC.  The high voltage feedthrough held the cathode plane and voltage divider chain at $-18$\,kV. A Matsusada linear power supply provided a voltage from $-14$\,kV to $-19$\,kV, with currents around 20\,mA. Oscilloscope traces showed peak-to-peak variation of 10\,mV in the HV.   
A drawing and close-up image of the chemically etched grid on the cathode plane is shown in Figure~\ref{fig:cathode_plane}.

\subsubsection{Wire Planes}
The stack of wire planes consists of a so-called `grid' wire plane and three `sensing' wire planes that produce bipolar induction signals as ionized electrons pass by.  All three sensing wire planes are identical in their construction and framed by laminated copper clad FR4 receiver boards.  The gold-plated copper beryllium wires, ranging from 0.5\,m to 1\,m in length, were wound with a target tension of 130~gram-force. This tension was verified by mechanical oscillations in each wire induced by pulsing current through wires in the presence of a magnetic field \cite{bib:Convery}. Measured tensions were between 90 to 130 gram-force. The wire-mesh ground plane sits above the stack.  The grid plane defines the electric field in the sensitive drift volume for the u-plane. Specifically, the grid plane masks the u-plane from the field of the approaching ionization electrons. This ensures that the electrons pass the u-plane and the following v-plane while inducing the same bipolar peaks on both planes. Additionally, the grid plane enables clear determination of the midpoint of each bipolar signal, and eases hit-finding on any channels exhibiting heightened noise levels. 

The grid, u-, and v-planes are held at bias voltages (relative to ground) of $-420$\,V, $-230$\,V and 0\,V, respectively, while the collection x-plane was held at $+270$\,V. The ground plane was maintained at the high-voltage ground. Bias power to the planes is delivered directly to the receiver boards. This power is distributed to the wires through a resistor chain with capacitive coupling to the readout connector.  The bias voltage is carried to each grid, wire and ground plane from the feedthrough boards using teflon coated wires.
The DC component is decoupled with a capacitor-resistor pair just after the wire attachment point.  The receiver boards have 72-pin motherboard connectors. 

\begin{figure}[!b]
	\centering
        \includegraphics[width=0.47\textwidth]{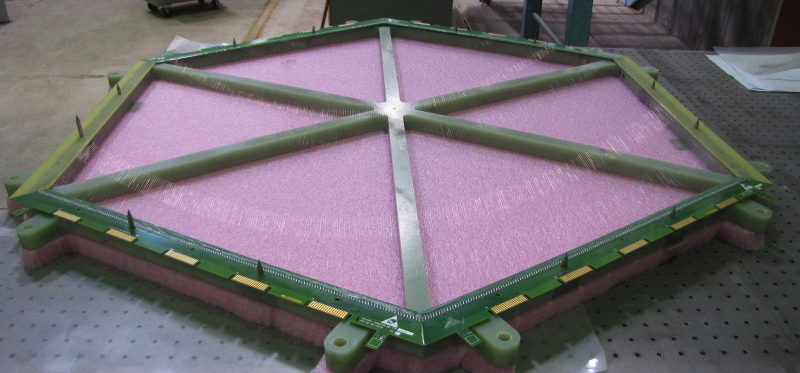}
        \caption{Image of a sense wire plane with copper beryllium wires spaced 3~mm apart on an FR4 frame.} 
        \label{fig:sense_wire_plane}
\end{figure}

\subsection{LArTPC Readout Electronics}
\label{sec:readout}
\begin{figure}[!t]
        \includegraphics[width=\linewidth]{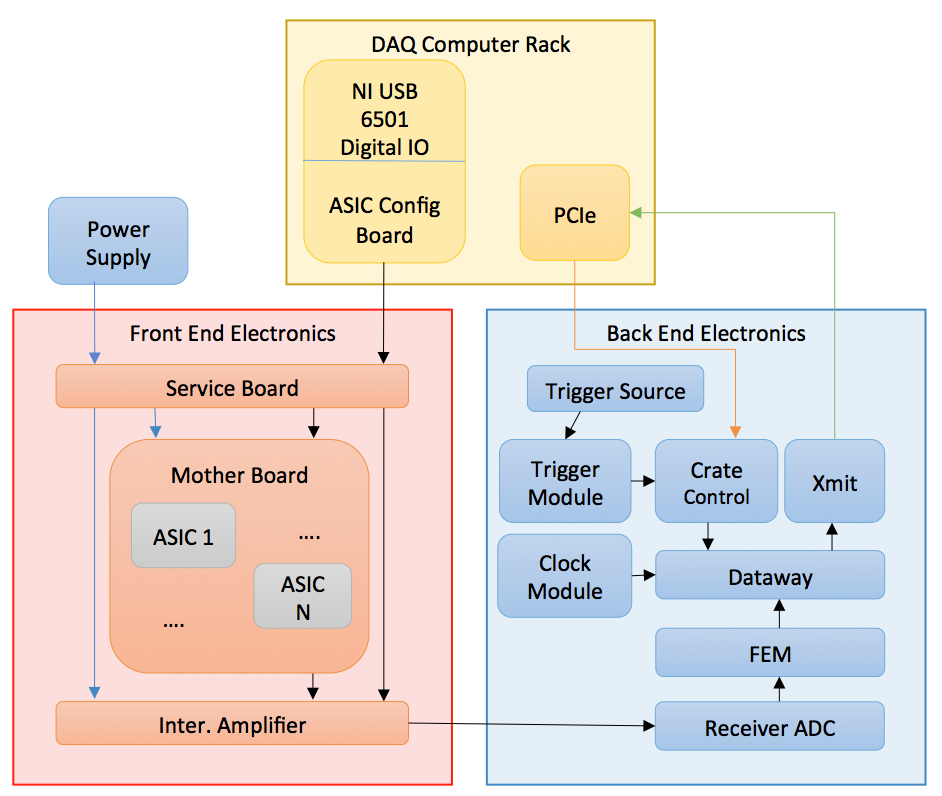}
        \caption{Flow diagram of the LArTPC readout electronics.} 
        \label{fig:electronics_diagram}
\end{figure}

The readout electronics is made up of three subsystems: the front-end electronics (FEE)~\cite{bib:Chen}, back-end electronics (BEE)~\cite{bib:Cheng}, and wire-plane data acquisition (DAQ)~\cite{bib:microBooNE}.  The word ``front-end'' denotes the electronics mounted directly on or inside the cryostat, e.g. the \textit{in situ} motherboards sitting in the liquid argon.  The word ``back-end'' denotes the triggering and signal processing systems sitting outside of the cryostat. A flow diagram of this system is shown in Figure~\ref{fig:electronics_diagram}.  

An application-specific integrated circuit (ASIC) configuration board is used to communicate instructions to the cold motherboards.  They consist of three digital input/output channels. A USB type-B cable delivers computer instructions to an off-the-shelf I/O board (NI USB6501), which is driven by the ASIC configuration board.  The digital serial configuration stream is delivered to the service card through a differential HDMI cable to the service board.  Apart from ASIC initialization, the systems are independently operable.

To initialize the BEE, the DAQ computer uses the ASIC configuration board to issue an ASIC rise-time and gain configuration instruction to the FEE service card. The service card relays that signal to the ASICs, along with power. Once initialized, the motherboards maintain continuous collection of signals from the wire planes and forward those signals to intermediate amplifiers.

The primary responsibility of the BEE is the uptake of analog signals from the wire planes via the receiver ADC boards. After analog-to-digital (ADC) conversion, these signals are passed to the front-end module (FEM) card for processing. The FEM card is controlled by a clock, thus continuously sampling and recording the ADC output even in the absence of a signal. Without a trigger, however, these data are dumped every cycle. On receipt of a trigger via the crate controller, a predetermined sample data size is taken from the FEM memory and passed to the transmit (XMIT) card.  The DAQ computer then pulls the data from the XMIT card.  The crate controller can be triggered from both the DAQ computer and the trigger module.   

\subsubsection{Front-End Electronics}
Reliable and consistent data-taking in a LArTPC requires minimization of electronic noise.  A thorough study of electronic noise in LArTPCs can be found in Reference~\cite{bib:Radeka}. The dominant forms of electronic noise come from first transistor noise, occurring during signal amplification process, and from thermal noise along the sensing wires and connection leads.  These forms of noise can be mitigated by minimizing the total capacitance of the sensing wires, signal cables, and input transistors. Cold front-end electronics, developed  by Dr. Veljko Radeka and Dr. Hucheng Chen at Brookhaven National Lab (BNL), operate with a minimal path length between the signal wires and the preamplifier thereby reducing the total capacitance seen at the preamplifier input.

Thermal noise from the sense wires was mitigated by the use of copper beryllium material. The thermal noise was calculated based on the transmission line model of the sense wires as described in Reference~\cite{bib:Radeka} and was not a noticeable problem during data-taking.  

Three large conflat flanges are instrumented atop the cryostat.  Two of these flanges are used for front-end electronics feedthrough. Of all the FEE components, only the motherboards sit inside the liquid argon environment, connected directly to the termination/receiver boards on the wire planes. Cold cables run from the motherboards out to ``feedthrough boards" atop the corresponding conflats.  These receive power and configuration information from the service board, and return the wire-plane signals. The remaining FEE components sit inside Faraday boxes on each of the feedthrough circuit boards.

Each feedthrough board supports two columns of line driver cards, with each column managed by one service board (developed by BNL).  Bus boards run along the side of the service board and intermediate amplifiers (line drivers), supplying power. The service boards supply approximately $\pm1.8$\,V to the front end ASIC chips on the motherboards, and can also supply calibration pulses.  Each board contains 1 molex power connector, 2 USB (type-A) ports for injecting the calibration pulse, and 2 HDMI configuration ports. An Agilent E3648A supplied  $\pm 3.3$\,V to the power connector.

\begin{figure}[!tb]
	\centering
        \includegraphics[width=0.48\textwidth]{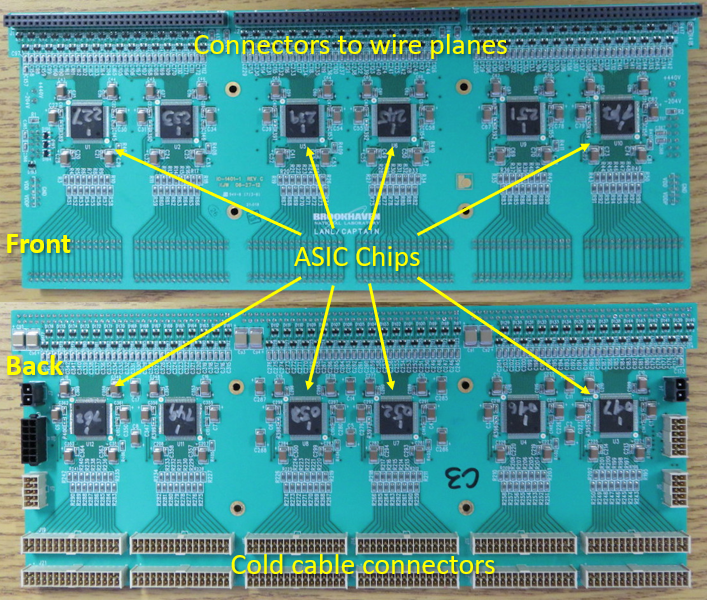}
        \caption{An image showing the front and back sides of a BNL cold motherboard. 
        }
        \label{fig:motherboard}
\end{figure}

The cold motherboards, also developed by BNL, are designed to operate in liquid argon.  Each motherboard utilizes 12 ASIC chips (six per side), and each ASIC chip reads out 64 TPC wires. An image of the front and back side of the cold motherboard is shown in Figure \ref{fig:motherboard}. 
The primary purpose of the motherboards is to amplify and shape the wire output signal. Data-taking runs typically use a gain of 14.7\,mV/fC for this purpose. 
The motherboards also pass calibration signals and injection pulses from the service card input to the ASIC chips.  Optionally, bias can be supplied to the wires through the motherboard. However, the standard mode of operation is to apply bias directly to the wire planes.


Upon receiving signals from the motherboards, intermediate amplifiers split them differentially and apply a gain of 12\,dB. This ensures clean transmission of the analog signal to the BEE located 10\,m away from the cryostat. A 68-pin SCSI-3 connector carries the differential signal from each of the 32 channel amplifier cards.

\subsection{Back-End Electronics}
The BEE consists of TPC VME crates, clock module, PDS VME crate, calibration fanout card, and power supplies~\cite{bib:Cheng}. The TPC VME crates were assembled and tested at Nevis Labs, at Columbia University. Each crate consists of TPC readout boards (mechanical assembly), crate controller, trigger module, XMIT card, dataway back-plane, and fanout card.  Every crate is controlled and read out by a DAQ sub-event builder (SEB) computer, which passes data onto the main assembler computer. 

\begin{figure}[!htb]
	\centering
        \includegraphics[width=0.48\textwidth]{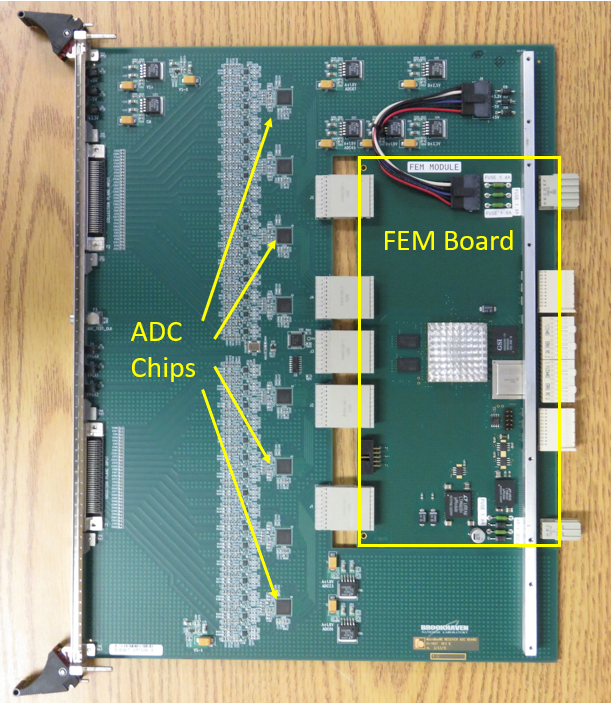}
        \vspace{1mm}
        \caption{The mechanical assembly of the TPC readout board showing how the BNL and Nevis boards were joined together. The FEM daughter card fit into the ADC-receiver board. }
        \label{fig:mechanical_assembly}
\end{figure}

Each TPC readout board receives data via two 32-channel differential warm cables. The board incorporates two sections into a VME 9U board: the digitizing section designed and built at Brookhaven National Lab, and the data handling section built by Nevis Labs. The digitizing section is laid out as an 8-layer printed circuit board.  The ADC receiver board is a 14-layer printed circuit board, with an interface to the daughter Nevis front-end module (FEM) board as shown in Figure~\ref{fig:mechanical_assembly}.  More detail on the boards can be found in the following section.

\subsubsection{ADC Receiver Board}
The ADC receiver board incorporates 8-channel ADC chips (AD9222) and line receivers. The ADCs sample at a rate of 2 MHz which is the optimum choice of the sampling frequency given the corresponding time scale of the electron drift velocity, wire plane spacing, and diffusion~\cite{bib:uboone_noise}. With low power consumption (less than 100 mW per channel) the AD9222 generates high integration density for the readout board. It also creates a high-speed low-voltage differential signal (LVDS) serial output stream.  This reduced the number of FPGA connections required. To accommodate the various signal properties on the collection and induction planes, the collection plane signal baseline was made adjustable. The adjustments also ensure that the collection plane's unipolar differential signals could make use of the full ADC analog input range.

Two SCSI-3 connectors on each receiver board accept signals from 64 TPC wires.  The receiver boards use four HM-Zd connectors to send the differential outputs to the FPGA chip.  These connectors had individual ground shielding on each pair. The PDS triggers, clock signals, SPI signals, and light emitting diode (LED) signals were sent through a fifth HM-Zd connector.  The receiver boards also includes a daughter FEM board on its backend described in the following section.
\vspace{1mm}

\subsubsection{FEM Board}
\begin{figure}[t!]
	\centering
        \includegraphics[width=0.48\textwidth]{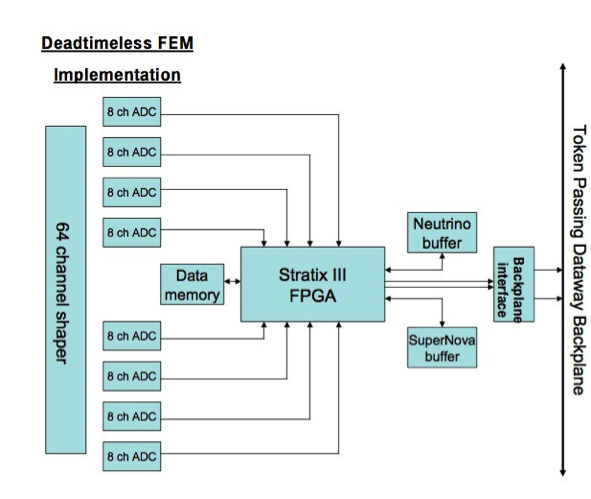}
        \caption{The digital readout scheme showing the 8 ADC chips, the FPGA, and the memory chips. Illustration taken from Ref.~\cite{bib:microBooNE}.}
        \label{fig:digital_readout}
\end{figure}

The data-processing chain begins with the front-end module (FEM) boards. Each FEM board includes a Stratix III Altera FPGA \cite{bib:microBooNE}, which accepts the ADC-reciever board's digitized outputs as shown in Fig. \ref{fig:digital_readout}. Each FPGA uses a 1M $\times$ 36 bit 128 MHz SRAM memory to save data from 64 wires sequentially in time. The data is grouped into two 36 bit ADC words. This requires a 64 MHz data storage rate ($64/2 \times 2$ MHz ADC sampling). The speed and size of the SRAM memory allows continuous readout of the TPC data.  

The data-processing chain was configured to allow high uptake of data that coincide with beam-induced neutron interactions in the liquid argon.  Accelerator beam bunches containing an interaction in the TPC drift volume were identified using the PDS to detect a prompt flash of light.  A data frame header contained the beginning frame marker, the event type, the trigger position (if present), the channel address, and the word count.  The readout of the clock is not synchronous with the neutron beam spill time.  Consequently, each 4.8\,ms window of relevant event data spans four 1.6\,ms frames. To reduce the data rate, the FPGA trims the four frames to span the exact 4.8\,ms required, with 1.6\,ms being sampled before an event trigger and 3.2\,ms afterward.

\subsubsection{Trigger Board}
The trigger board receives booster and WNR beam gate signals.  The beam gate signals can be used to flag PDS triggers as neutron interaction candidates.  PDS triggers that are valid can be compared with a logical OR operation to utility triggers received by the trigger board thereby outputting calibration triggers.  These calibration triggers can use various measurements including random triggers for noise measurements, off-beam triggers for cosmic ray response studies, and triggers for electronics studies.  

\subsubsection{Clock Module and Fan-out Cards}
The clock module sends a low jitter signal from a 16 MHz clock to the two VME crates with ADC receiver boards~\cite{bib:microBooNE}. The clock fan-out on the back of each crate distributed this timing signal to each TPC readout board. A local 16 MHz crystal on the crate clock fanout board generates a local clock for event gating (pretriggering). 

The clock module was found to be sensitive to both the power source and ground. When the clock module occupied a NIM crate shared by other modules, the DAQ system would crash after a few hundreds of events. Isolation of the clock module into its own crate, with separate power and ground, reduces the frequency of DAQ crashes.

\subsection{Feedthroughs}
\label{sec:feedthroughs}
The signal transmission between the TPC and the readout electronics system consists of cold cables, warm cables, and signal feedthroughs. Gold plating on the exterior of the feedthrough board serve as the ground for the signals and power.  The cold cables transmit detector signals from the cold motherboards to the intermediate amplifiers and also distribute power to analog front-end ASIC chips.  The cold cable is a custom-built 32 twisted pair flat cable with Teflon FEP insulation. It used an AWG 26 solid core and silver plated conductors providing approximately \SI{100}{\ohm} impedance. The 64-pin termination uses 0.100"-pitch dual-row crimping. Correct alignment with the signal feedthrough pin carriers is ensured with custom shells and jack screws. 

\begin{figure}[!b]
    \centering
        \includegraphics[width=.45\textwidth]{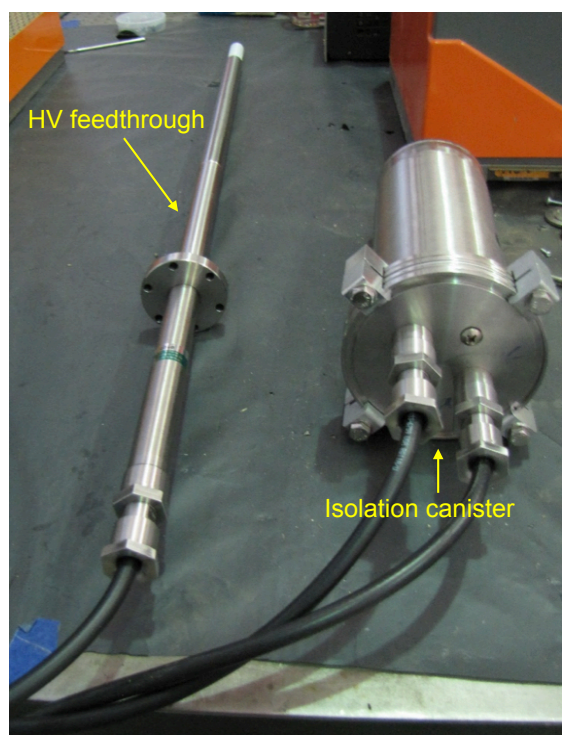}
        \caption{The HV feedthrough assembly with cable is shown on the left and the electrical isolation canister is shown on the right.}
        \label{fig:HV_feedthrough}
\end{figure}

Signals between the intermediate amplifier and the receiver/ADC board on the readout crates are transmitted through 68-pin SCSI-3 warm cables, each with 34 differential pairs and two MicroDensity connectors. Shielding was fashioned from aluminum foil with 10\% overlap.  

The Faraday boxes mounted on top of the feedthrough boards shielded the intermediate amplifiers. One box also holds the four bias voltage feedthroughs, providing -430, -230, 0, and +230\,V to the detector wire planes.

HV feedthrough enters the cryostat on a dedicated 2 3/4" conflat flange. The cable's own capacitance and a 25\,M$\Omega$ resistor provides RC filtering. Figure~\ref{fig:HV_feedthrough} shows the canister used to shield the resistor, assembled with a shielded feedthrough extension to deliver HV to the bottom of the TPC. An isolation standoff separates the flange from feedthrough, but the canister was eventually grounded to the cryostat during stable data-taking.

\subsection{Photon Detection System} \label{sec:pds_config}
\begin{figure*}[t!]
    \centering
    \begin{subfigure}[t]{0.48\textwidth}
        \centering
        \includegraphics[height=2in]{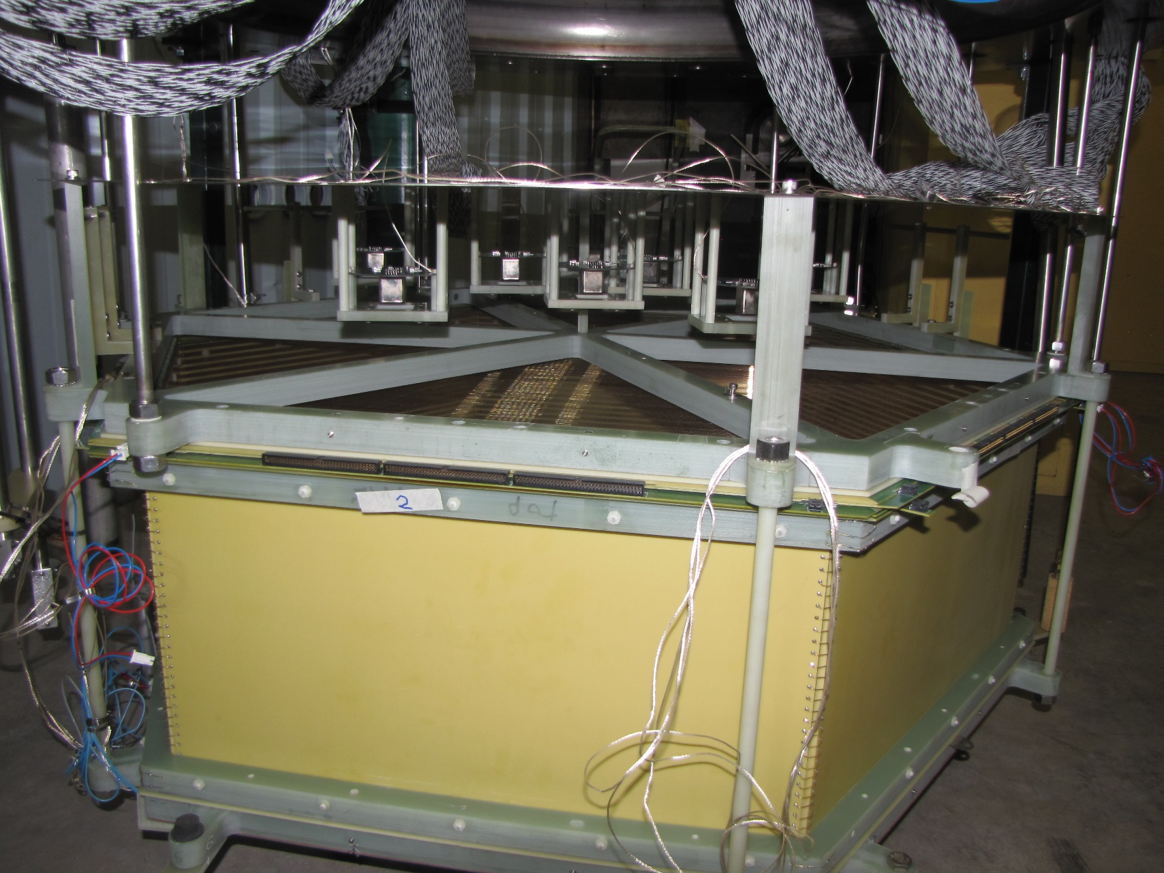}
        \caption{}
        \label{fig:pmt1}
    \end{subfigure}%
    \begin{subfigure}[t]{0.48\textwidth}
        \centering
        \includegraphics[height=2in]{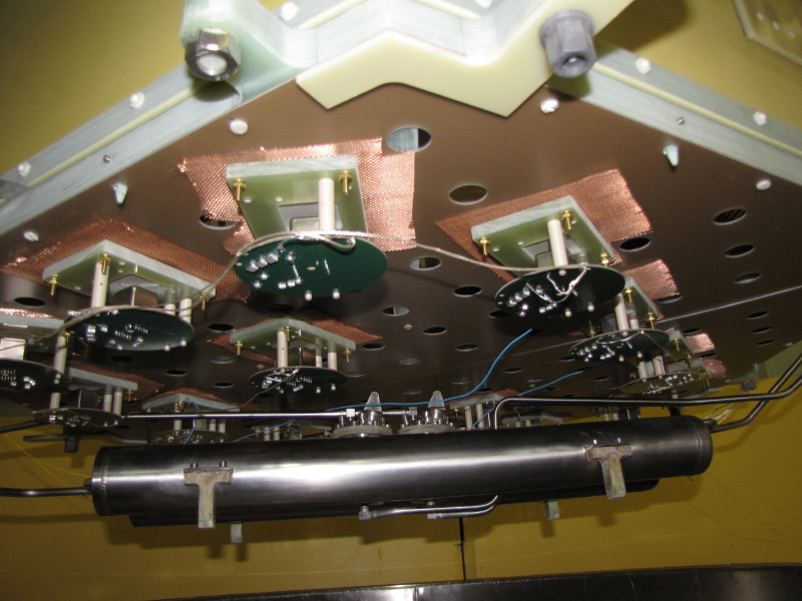}
        \caption{}
        \label{fig:pmt2}
    \end{subfigure}%
    \vspace{0.5cm}
    \begin{subfigure}[t]{0.48\textwidth}
        \centering
        \includegraphics[height=2in]{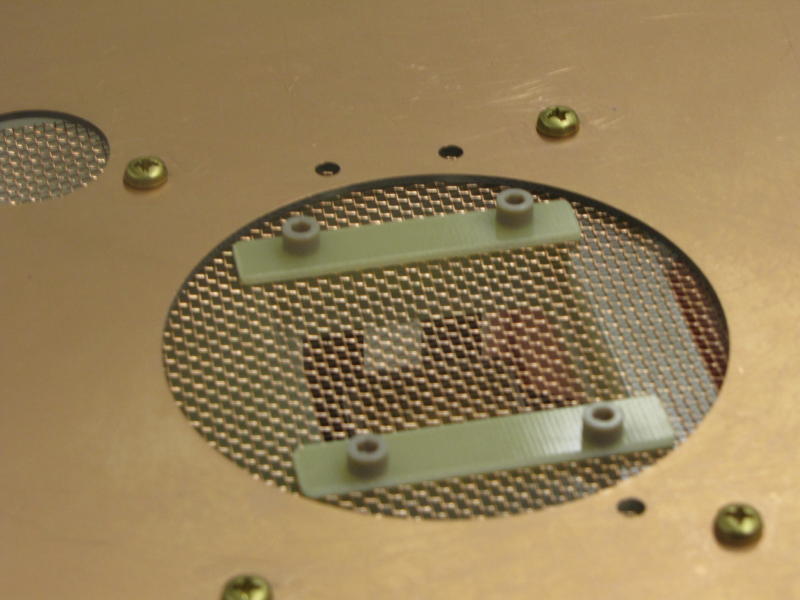}
        \caption{}
    \end{subfigure}%
    \begin{subfigure}[t]{0.48\textwidth}
        \centering
        \includegraphics[height=2in]{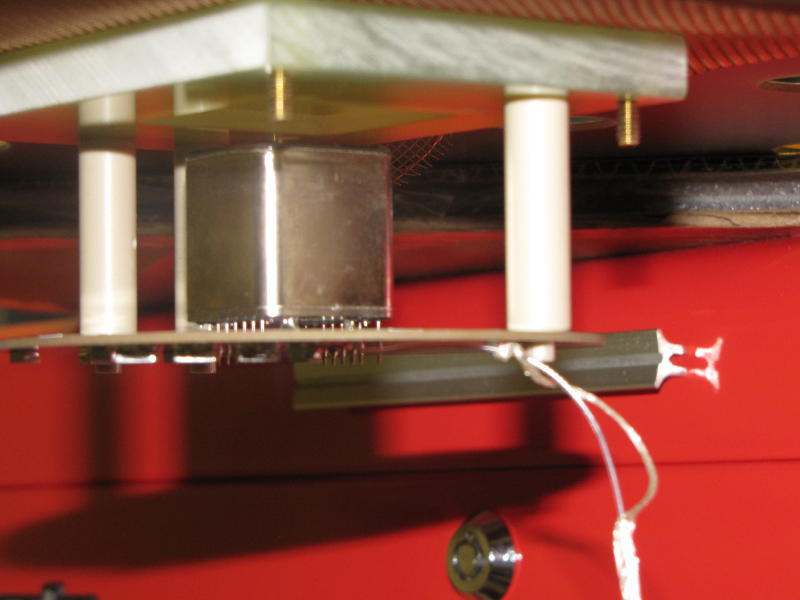}
        \caption{}
    \end{subfigure}
    \caption{Images (a) and (b) show the locations of the PMTs with their bases mounted above the sense wire planes and on the bottom of a copper plate, respectively. Image (c) shows the  copper grid just beneath the copper-clad FR4 plane and (d) shows a PMT and base sitting just beneath a TPB coated window.}
        \label{fig:pmt_config}
\end{figure*}

The PDS provides crucial improvements to neutron candidate event timing and energy reconstruction beyond what can be achieved by the LArTPC alone.
Typical readout times for LArTPCs are on the order of one millisecond, whereas LAr scintillation light detection enables nanosecond timing. 
The projected nanosecond timing resolution is critical to allow separation of the prompt gamma-ray backgrounds from the subluminal neutrons during data-taking in the neutron beam. 
Furthermore, nanosecond timing resolution from the PDS allows the incident neutron energy spectrum to be reconstructed using time-of-flight. 

\subsubsection{Photomultiplier Tubes}

The PDS is made up of twenty-four Hamamatsu R8520-506 MOD PMTs. 
The R8520 is a compact rectangular PMT which is approximately 1" $\times$ 1" $\times$ 1" in size. A borosilicate glass window exposes a special bialkali photocathode capable of operating at liquid argon temperatures (87\,K). 
The R8520-506 MOD's peak quantum efficiency is 25\% at 340\,nm. Liquid argon scintillates at a VUV wavelength of 128 nm, which would be absorbed by the borosilicate PMT window before reaching the photocathode. 
Therefore, the VUV photons must be shifted towards the visible wavelengths before interacting in the PMT glass.  This is accomplished positioning TPB-coated acrylic windows in front of the PMTs to shift VUV scintillation light into the visible wavelengths. 
Tetraphenyl butadiene (TPB) is the most commonly used wavelength shifter for liquid argon detectors and has a conversion efficiency of about 50\% when evaporated in a thin film ~\cite{bib:Benson}. 
The re-emission spectrum of TPB peaks around 420\,nm~\cite{bib:Gehman}. 
 
The PMTs are split into an array of 8 PMTs on the top plane and an array of 16 PMTs bottom plane. 
The top plane is mounted on a frame just above the sense wire planes. 
The TPC ground plane, just above the sense wires, shield the PMTs from any residual TPC electric field. 
The bottom PMT plane is fixed behind the TPC cathode plane, with a copper-clad FR4 plane for electric field shielding. 
A copper grid sits between the copper-clad FR4 plane and the TPB coated window in front of each PMT. The positioning of PMTs relative to the TPC is shown in Figure~\ref{fig:pmt_config}.

\subsubsection{Photon Detection Electronics}

\begin{figure}
    \centering
    \includegraphics[width=\columnwidth]{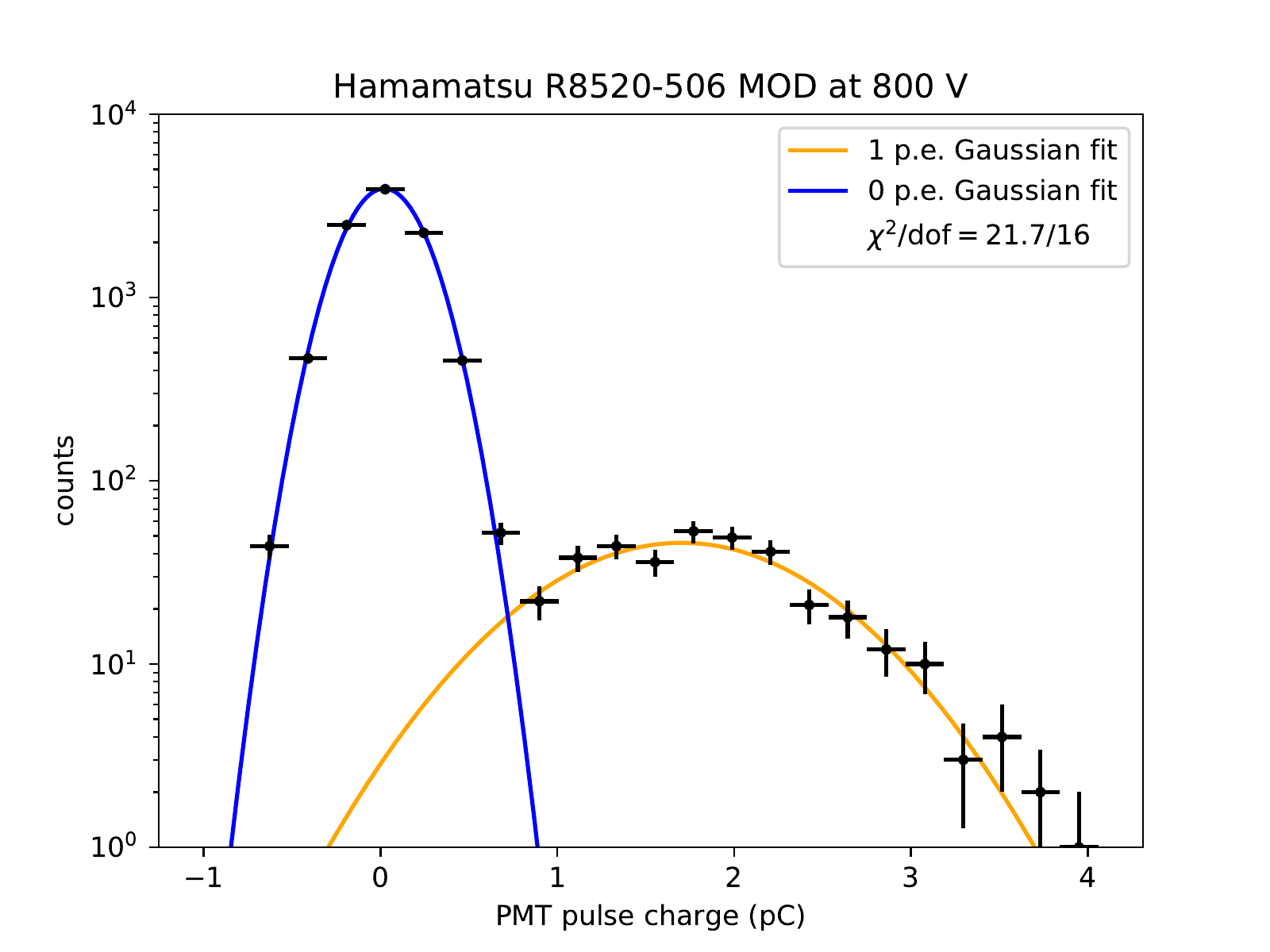}
    \caption{Typical PMT charge distribution measured at showing the single-photoelectron response at a bias of 800\,V and less than 5\% occupancy.}
    \label{fig:pmt_charge}
\end{figure}

One Bertan 205A-03R supply powers all 24 of the PMTs. 
A positive bias of 800\,V is daisy chained between three voltage distribution boxes and passed to signal pickoff boxes developed at UC Davis.
The PMT bases were jointly designed by LANL and UC Davis. 
The pickoff boxes supplied power to each PMT base and decoupled the signal to the CAEN digitizers. 
The controller card passed commands from the DAQ to the digitizers, which triggered and recorded the signals. A typical single photoelectron charge distribution for a bias of 800\,V is shown in Figure~\ref{fig:pmt_charge}.  The data was collected when the PMTs had less than 5\% occupancy, and a peak-to-valley ratio of $\sim$2.5 was observed.

Approximately 600\,V was supplied to the PMTs during initial light leak checks after the cryostat was sealed shut. 
The largest source of light inside the cryostat was found to be the electrical and thermal isolation standoff on the top head. 
Significant light leaks were also found in all SHV voltage feedthroughs not fitted with a connector.
Plastic and electrical tape were used to optically seal any suspected leak-points, and a removable cover was used to occlude the laser viewport. 
The PMT operating voltage was increased towards optimal levels as the light leaks were mitigated.

\subsubsection{Photon Detection DAQ}

PDS signals are digitized by three CAEN V1720 digitizers. 
The V1720 is an 8-channel, VME-based digitizer that samples at 250 MS per second (4\,ns) at 12 bits. 
While Mini-CAPTAIN deployed 24 PMTs, only 21 PMTs are digitized in order to accommodate the additional RF synchronization pulse as input to each of the three V1720 digitizers. 
During initial testing, a VME crate controller was used to collect the data from the VME backplane and transmit to a DAQ computer via a single optical fiber uplink. 
This system was upgraded to an individual optical link between each digitizer board and the computer to greatly improve the data throughput for high-rate running conditions. 

For the majority of data taking in the neutron beam, the digitizers were triggered externally. 
The first beam pulse forced a zero-bias external trigger in both the PDS and TPC in order to maintain sub-system synchronization. 
After the first forced trigger, subsequent PDS triggers within the TPC acquisition window (${\sim}4$\,ms, to identify cosmic ray backgrounds) were driven by PDS activity. 
Each PMT channel had a threshold level set to ${\sim}0.25$\,photoelectrons. 
The V1720 has a free running level check that reports the number of threshold crossings on its analog front panel output: a voltage signal of 125\,mV $\times$ [number of level crossings] is generated. 
These signals are fanned together in NIM-based analog electronics where a level discriminator selected the multiplicity of simultaneous PMT hits. 
For the neutron beam data taking runs, 4 PMT hits are required to trigger TPC data acquisition.

Approximately 8\,$\mu$s of digitized data was collected per trigger, which sufficed to extract nearly all of the singlet and triplet light. 
Data was read from the digitizer boards and stored in shared memory. 
A second, asynchronous program pulled these data from shared memory and reconstructed the events in a ROOT tree~\cite{bib:root}. 
Barring communication loss between the computer and digitizers, this created a dead-time free system. 

The digitizer clocks for each board were not disciplined to an external oscillator because of the short duration of the neutron run. 
Instead of a GPS, the TPC and PDS computers were synchronized through an NTP timeserver. 
This provided ${\sim}10$\,ms timing synchronization between the PDS and TPC triggers, which was sufficient given that a 200~ms hold-off was used between beam triggers. 
The relative alignment of each digitizer clock does drift over the course of a run,
but the alignment of the digitizer clocks is easily achieved with common event spacing, trigger multiplicity, and the presence of the RF synchronization pulse.  A GPS module was used to assign a time-stamp to each PDS event. 

\subsection{Power Distribution and LED Testing}

\begin{figure*}[t!]
    \centering
    \begin{subfigure}[t]{0.48\textwidth}
        \centering
        \includegraphics[height=2in]{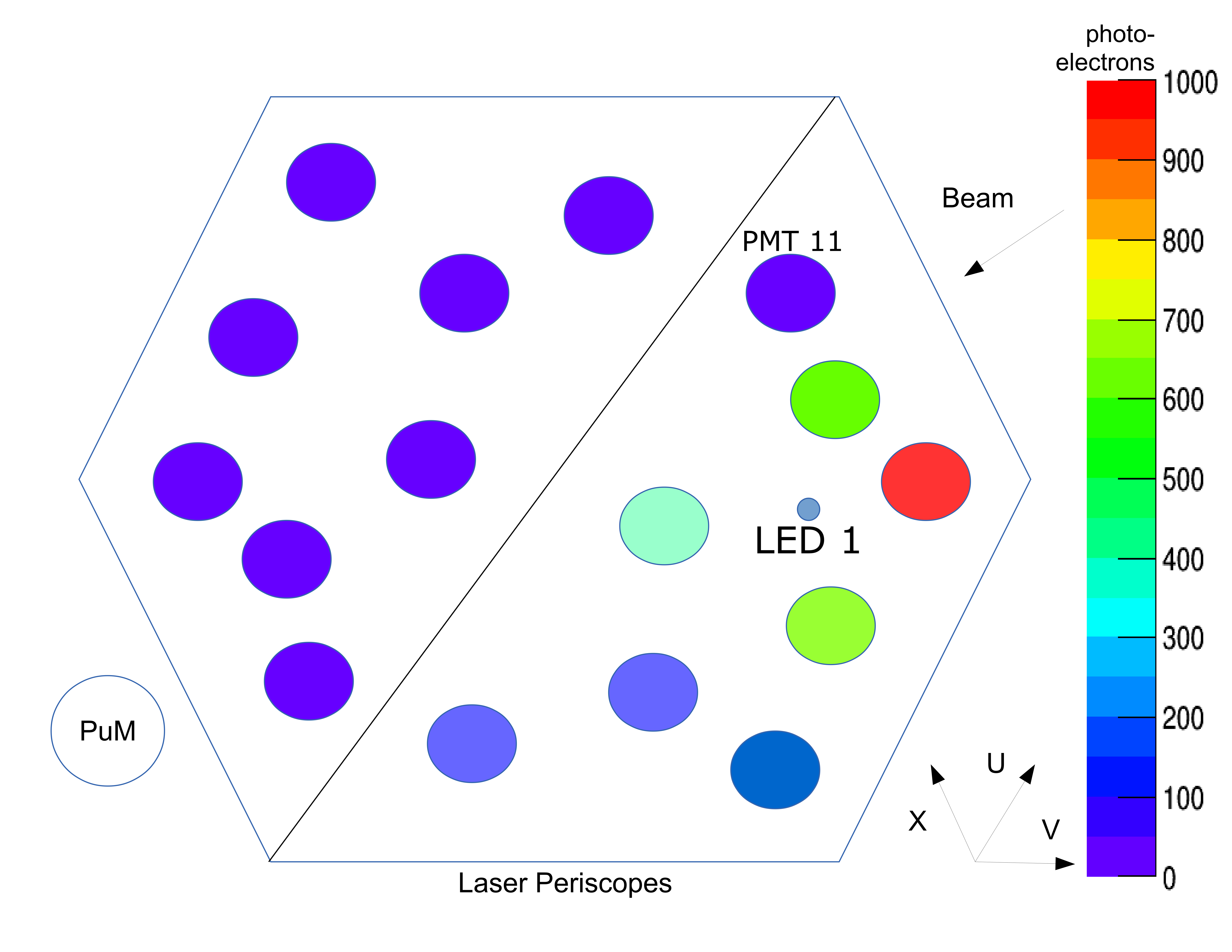}
        \caption{}
    \end{subfigure}%
    \begin{subfigure}[t]{0.48\textwidth}
        \centering
        \includegraphics[height=2in]{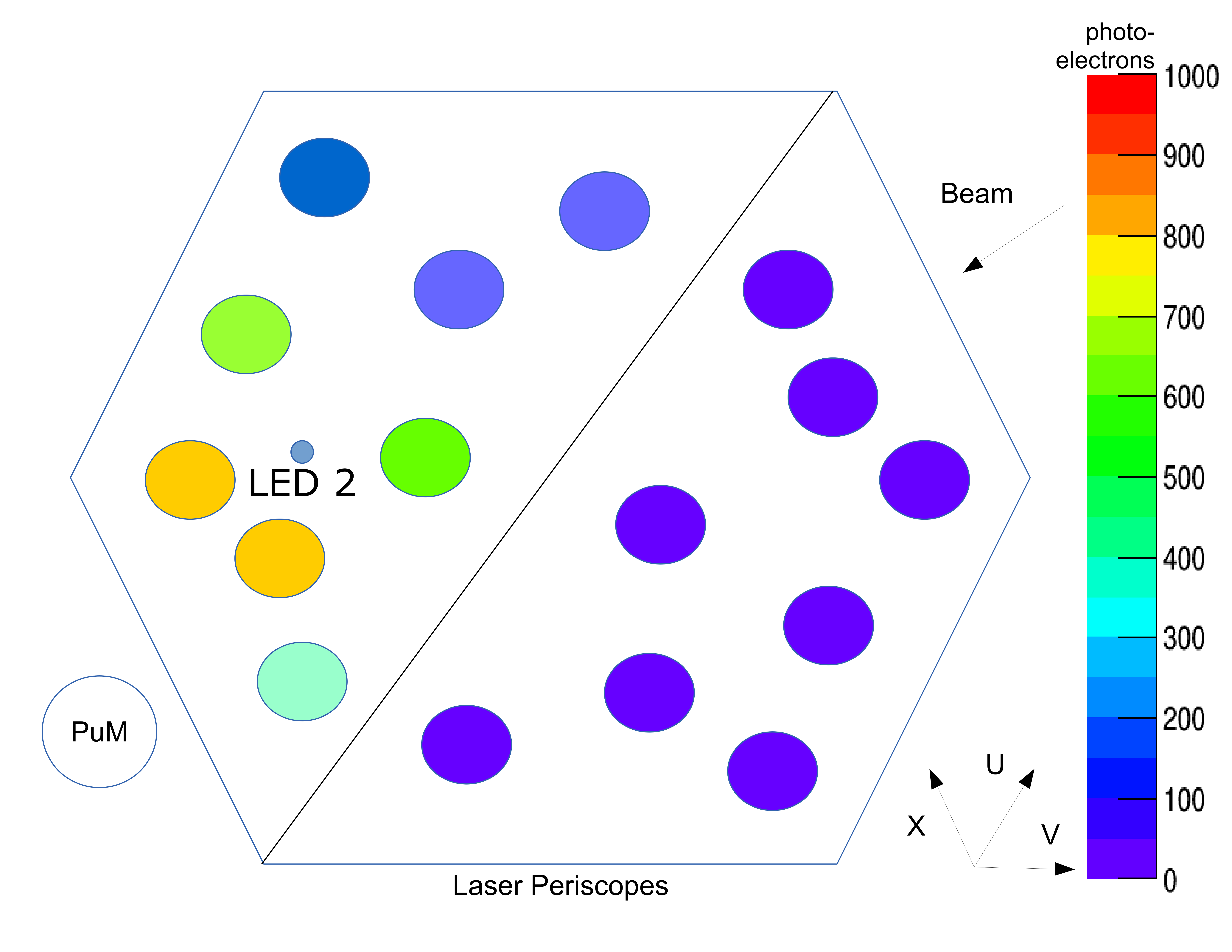}
        \caption{}
    \end{subfigure}
    \caption{The PMT configuration for the PDS system, and the system's response to a calibration LED mounted on the (a) right or (b) left side of the field cage (looking downward).}
        \label{fig:pmt_config_02}
\end{figure*}

LEDs are installed on each end of the field cage to enable testing of the PMTs during the purification of the liquid argon. 
Figure \ref{fig:pmt_config_02} shows the results from the initial PDS test before moving Mini-CAPTAIN to the WNR neutron beam facility. 
There was some initial concern with the response on the 11th PMT, but later tests showed the PMT yielded signals comparable to those from the other tubes in response to neutron events. 
It is likely that the TPC support frame shadowed the direct line of sight from the calibration LED to PMT 11 as can be seen in  Figure~\ref{fig:pmt_config_02}(a).  

The LED system is also used to calibrate the single photoelectron response for each PMT. 
Since the PMTs are biased with positive high voltage, the photocathodes in liquid argon are at the ground potential. 
The readout, sitting outside of the cryostat, requires capacitive coupling, which introduces slight overshoots and ringing. 
LED calibrations are used to measure the impulse response of the PMT and later correct for these artifacts.

\section{Mini-CAPTAIN Commissioning} \label{sec:miniCAPTAIN_comm}
\begin{figure*}[t!]
    \centering
    \begin{subfigure}[t]{0.48\textwidth}
        \centering
        \includegraphics[height=2.02in]{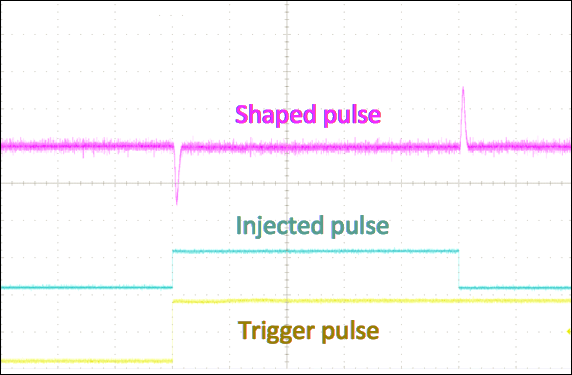}
        \caption{}
    \end{subfigure}%
    \begin{subfigure}[t]{0.48\textwidth}
        \centering
        \includegraphics[trim={0 1cm 0 1cm},clip, height=2.1in]{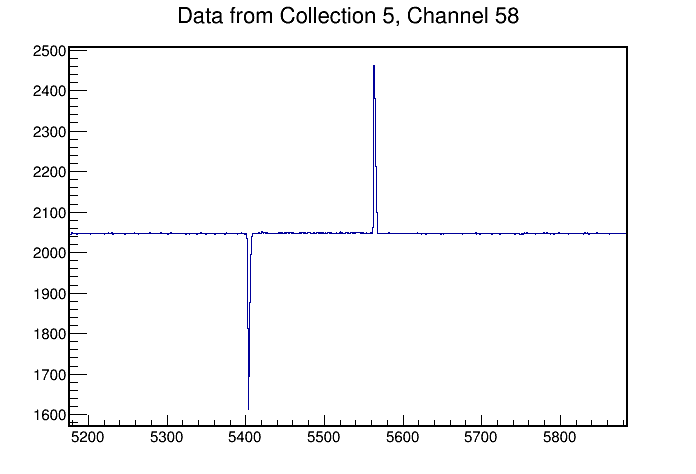}
        \caption{}
    \end{subfigure}
    \caption{The calibration injection pulse after the shaping capacitor for is shown as an oscilloscope trace, including the original pulse and trigger, in (a).  The final DAQ output of the calibration pulse is shown in (b).}
        \label{fig:injection_pulse}
\end{figure*}

The commissioning campaign can be divided into three stages: warm cryostat testing, liquid nitrogen testing, and liquid argon testing.  Initial warm tests only used the front-end electronics. 
During these initial tests, oscilloscope signals were primarily used.  Some tests were also performed using Nevis Labs FEM readout software~\cite{bib:Chen}. Before cold testing began, a complete end-to-end characterization of the electronics chain was made both in atmosphere and in vacuum. For these measurements, the high voltage was left off to prevent arcing.

\subsection{Power Isolation}   \label{sec:power_source}
Mini-CAPTAIN was staged in a building at the end of LANSCE Experimental Area A. The staging area had so-called clean and dirty sources of power.  Tests showed the clean power to be only slightly more stable when compared with the dirty power. Regardless, efforts were made to run sensitive electronics on clean power only. Additional isolation transformers were used whenever possible to suppress noise from the power source. The cryogenic and vacuum systems ran on dirty power and 3-phase power. The DAQ computers ran on dirty 120 V outlets. The front-end and back-end electronics racks were both isolated using separate transformers. To further increase the stability of the DAQ, the clock module was connected to its own clean power source.   

\subsection{Motherboard Operation}
During initial testing, most of the calibration tests used a 200\,mV injection pulse. However, some of the very first calibrations used $\sim$500\,mV pulse heights similar to the 600\,mV pulse heights used by Nevis Lab~\cite{bib:GCheng}. It was later determined that continuous use of the $\sim$500\,mV injection pulses was harmful to the motherboards.  This was evident every time the calibration testing initiated and a pair of motherboards on a wire plane would no longer configure. Reasonable caution was used during the electronics assembly and it seemed unlikely that the damage came from electrostatic discharge. To avoid further issues the 200\,mV maximum pulse height was established, and in most cases 80\,mV was used. To further reduce systematic uncertainty during calibrations only specifically trained individuals were allowed to perform the calibration tests. After these additional measures were taken no further motherboard malfunction was observed.   

\subsection{High Voltage Operation} \label{HVBreakdown}
Prior to the liquid argon fill, it was necessary to test the electronics with the HV applied to the TPC. At this point, the cryostat was under vacuum.  After the HV was turned on, and allowed to increase to -15\,kV, the system tripped. The anode plane motherboards no longer configured properly, and there was a clear loss of $~$20\% of the ASIC channels on the u- and v-planes. The most likely cause of these problems appeared to be HV discharge in the ground plane that in turn damaged the motherboards.  

The HV discharge also damaged some of the decoupling capacitors on the wire plane. Near the end of the liquid argon commissioning run, a significant increase in the number of channels with high noise was observed. After opening the cryostat the wire current was measured across the decoupled capacitors on the respective channels. Up to 30\% of the bias on the affected wires was leaking into the motherboards, with the worst being channel 593. The traces were cut for any wire with greater than 1\,V of leakage.

Despite the damaged channels and subsequent repairs from the breakdown, the noise levels appeared low before the initial fill of liquid argon during the final liquid argon testing (an RMS around 5 ADC). However, after the HV was turned on again correlated noise across several channels near channel 593 was observed.  The source and solution to this problem is addressed in the next section.  

To reduce this risk of HV breakdown, isolating material was installed between the side of the cryostat and the cathode feedthrough tip, which limited the voltage allowed during the pre-fill testing. The motherboards were not replaced until wires with bad capacitors were isolated. These safeguards proved adequate in the subsequent data-taking run.

\begin{figure*}[!htb]
    \centering
    \begin{subfigure}[t]{0.48\textwidth}
        \centering
        \includegraphics[height=2.2in]{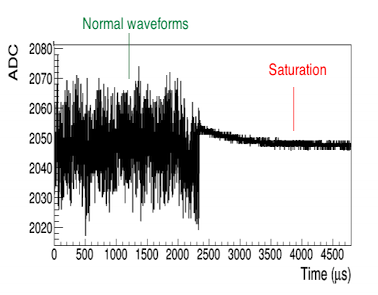}
        \caption{Beginning of saturation}
    \end{subfigure}%
    ~ 
    \begin{subfigure}[t]{0.48\textwidth}
        \centering
        \includegraphics[height=2.2in]{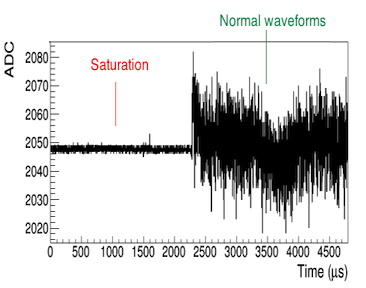}
        \caption{End of saturation}
    \end{subfigure}
    \caption{A normal waveform becoming saturated near the end of the 4.8\,$\mu$s acquisition window is shown in (a) and a saturated signal becoming unsaturated at the end of the window is shown in (b).}
\author[label1]{Q. Liu}
        \label{fig:saturation_image}
\end{figure*}

\subsection{Time Projection Chamber Testing} \label{sec:testResults}
Testing was performed in various TPC configurations, the most common configuration being an ASIC gain setting of 14 mV/FC, a rise time of 1 $\mu$s, and 500~pA of leakage current. The bias voltage was usually applied, while the high voltage was left off (even when the cryostat was filled with a cryogen). Most of the noise sources were associated with the front-end electronics, so it was most efficient to measure the outputs of the line drivers with an oscilloscope. Figure \ref{fig:injection_pulse} shows an example of the injection pulse used for testing. In this example, a 200 mV logic pulse with a width of 100 $\mu$s is used. The input capacitor alters the form of the square signal into bipolar peaks. The Nevis Labs FEM reads out the same signal to show the same bipolar peaks with a height of approximately $\pm400$ ADC units.  

Through all stages of the electronics testing, the baseline per channel remained consistent irrespective of the gain setting and FEM card used. Baselines also remained reasonably stable among channels within the same event.  Comparisons were made among events within the same run, as well as events from runs across different run cycles. Measurements with heavy noise can leave the impression of a baseline shift, depending on the sampling range and the form of the background signal. Smaller random pulses have the greatest effect on the code's sampling of the data. Some temporary shifts can be seen for a few milliseconds following a large noise spike. 

During the first phase of liquid argon testing, channel saturation was observed on some events. That is, a random channel on any of the wire planes would lose its signal and produce a baseline profile with a fluctuation of only a few ADC units. Figure \ref{fig:saturation_image} shows an event recording where the saturation starts and another event recording where the normal waveform returns. The exact duration of the lockout is unknown, because the start and end of a given saturation period were usually not simultaneously recorded. There were also cases in which only the flatline was observed within the window, leading to the conclusion the saturation time could be at least 4.8\,$\mu$s long.  

ASIC response tests were performed on the test stand at LANL. The ASIC chips were found to be fairly robust in their motherboard configuration. Pulses were directly applied to the leads on the 72-pin connectors on the motherboards, where they had to pass through the protection diodes. The pulses ranged from 20\,mV to 2.0\,V, with short 20\,$\mu$s and long 200\,ms widths. The pulse variation had no impact on the ASIC performance. However,  the act of touching a wire to the connector would make the ASIC trip for short time. This duration was dependent on how long the wire was in contact. For contact lasting longer than 2 seconds, the lockout was approximately 52 seconds. For a contact lasting lasting only a second or less, the channel would saturate for close to 10 seconds. No variation was seen with change of the ASIC configuration settings. A capacitor greater than 10\,pF in series prevented the lockouts on the test stand.  

A more complete study was performed at BNL~\cite{bib:HChen}. They found that this effect can be significantly reduced by configuring the ASIC current bias to 500\,pA. The saturation requires the positive current ($I_{\mathrm{pos}}$) to exceed the internal bias current and the product of $I_\mathrm{pos}\cdot\Delta t_\mathrm{pos}/C_{F}$. The saturation can be avoided if the ripple, given by $V_\mathrm{ripple}/\Delta{t}_\mathrm{pos}\approx I_\mathrm{pos}/C_{F}$, is less than 500 pA.  

The problem only persisted for a few weeks, so there was little motivation to continue studying this phenomenon. During the first run cycle (October 2014), the lockout was observed to happen on one or two channels every few events. This saturation was no longer observed after significant noise reduction efforts were completed. For this reason, the ASIC current bias was left at 100 pA.  

The wire planes became the primary focus for noise elimination after the commissioning run cycle was complete. The traces were cut for wires with significant leakage and damaged motherboards were replaced. The new motherboards were tested on the test stand prior to installation to ensure all ASIC channels were read out properly. Some noise was still evident during the first neutron run.  Prior to the second neutron measurement cycle, some additional traces were cut on the termination boards of the wire plane. The noise levels were sufficiently low for the remainder of the experiment.

\section{Performance in a Neutron Beam} \label{sec:miniCAPTAIN_neut}
\begin{figure*}[!ht]
	\centering
        \includegraphics[trim=3cm 0cm 3cm 1.5cm, clip,width=0.85\textwidth]{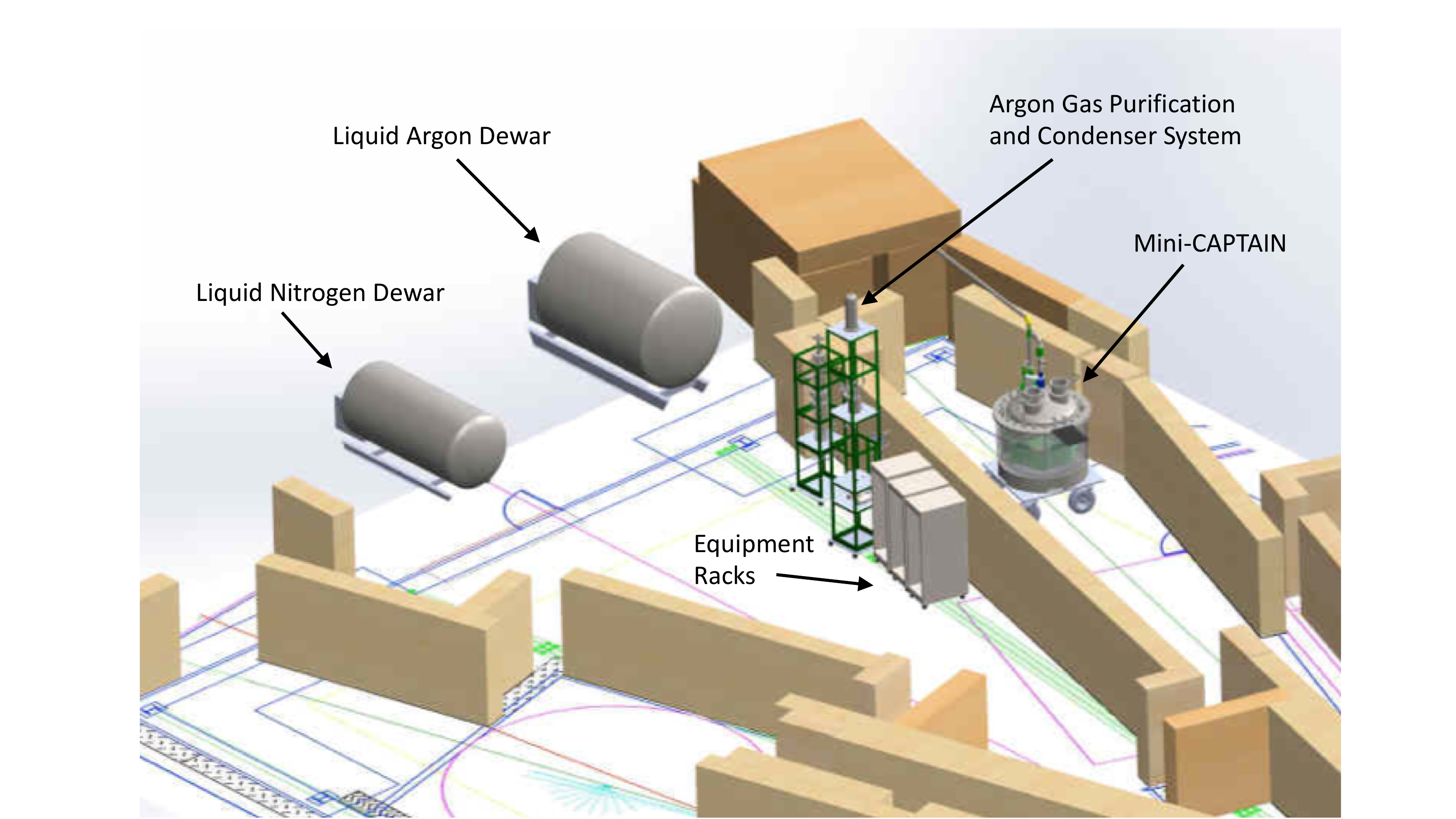}
        \caption{A diagram showing the experimental setup for neutron beam data-taking at the WNR facility, using the LANSCE accelerator.  The neutron beam originates from the lower right corner of the figure, directed towards Mini-CAPTAIN.}
        \label{fig:wnr_setup}
\end{figure*}

Mini-CAPTAIN measured neutron interactions in argon at the WNR facility at LANSCE, in a flight path known as 15R. The recirculation system, electronic racks, and computers were situated along the wall outside the beam path. Figure~\ref{fig:wnr_setup} shows the experimental setup within the WNR facility.  Due to the long purification time of the liquid argon, the cryostat was initially filled and purified outside of the flight path enclosure. A few days before our neutron run, two of the shielding blocks shifted, and the Mini-CAPTAIN detector was moved into position on top of a rolling platform. 

The first neutron run was performed in February 2016. It was clear from the start of data-taking that the purification system was not performing well.  The ionized charged tracks observed by the TPC were not robust enough for any meaningful physics studies.  However, this data proved useful for performance studies of the electronics.  The data collected was also useful for estimating the purity of the liquid argon, beyond the limits of the RTGs. Substantial improvement in the liquid argon purity was achieved by the second neutron measurement in May 2017.  

There were two phases of neutron data-taking consisting of high intensity and low intensity runs. For the high-intensity runs, the beam was delivered at its usual intensity of 3.8\,$\mu$A. The shutter to flight path 15R was partially closed to reduce the rate of neutrons interacting inside the detector. During the first few hundred runs, a high neutron rate (order 10 Hz) was observed and the shutter opening was reduced further until an event rate under one neutron per second was achieved. On the last day of data taking, the accelerator operated at a low intensity with the shutter open for a dedicated period of time to deliver a nominal 1\,Hz neutron interaction rate.

\begin{figure}[!htb]
	\centering
        \includegraphics[width=0.48\textwidth]{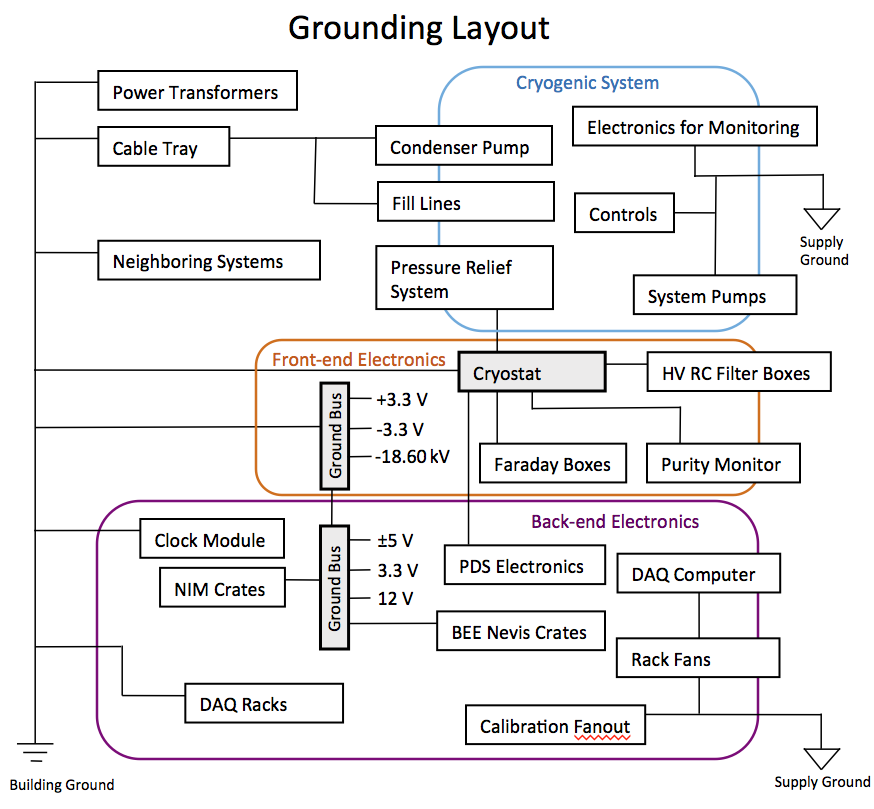}
        \caption{A flow diagram of the grounding layout for the experimental setup at the WNR facility.} 
        \label{fig:grounding_setup}
\end{figure}

\begin{figure}[!htb]
	\centering
        \includegraphics[width=0.48\textwidth]{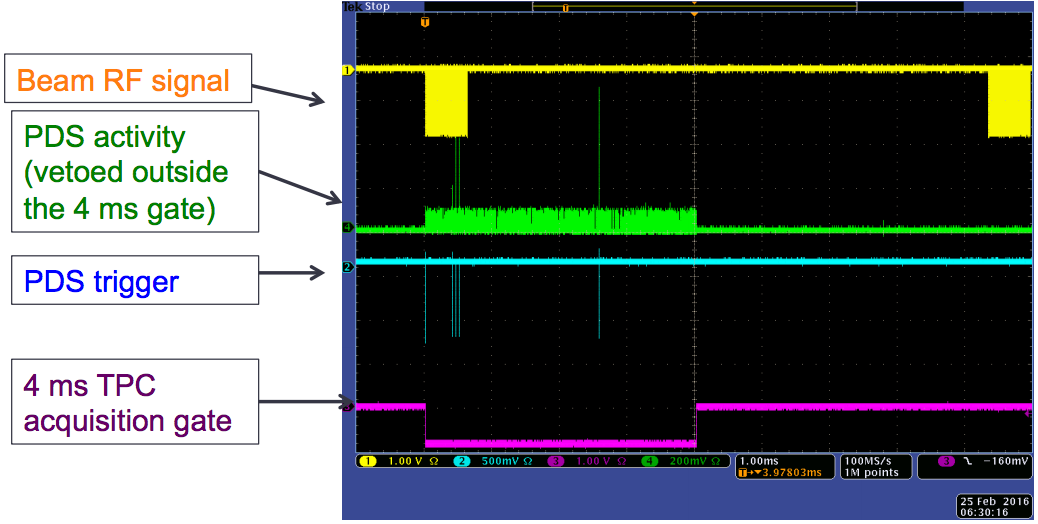}
        \caption{The traces of the beam RF signal, PDS activity, PDS trigger, and TPC acquisition gate as seen on the oscilloscope.}
        \label{fig:trig_oscope_01}
\end{figure}

\begin{figure*}[!htb]
    \centering
    \begin{subfigure}[t]{0.48\textwidth}
        \centering
        \includegraphics[height=1.8in]{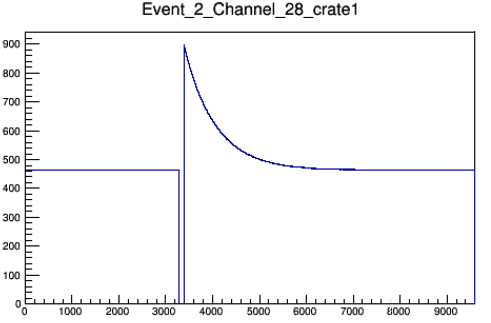}
        \caption{}
    \end{subfigure}%
    ~ 
    \begin{subfigure}[t]{0.48\textwidth}
        \centering
        \includegraphics[height=1.8in]{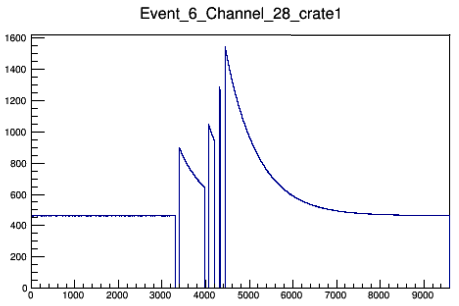}
        \caption{}
    \end{subfigure}
    \caption{TPC waveform readout of a single injected PDS trigger in shown in (a) and a similar event with four overlapping triggers is shown in (b).}
        \label{fig:PDS_trigger_in_TPC_output}
\end{figure*}

\begin{figure}[!htb]
    \centering
    \begin{subfigure}[t]{0.5\textwidth}
        \centering
        \includegraphics[height=2in]{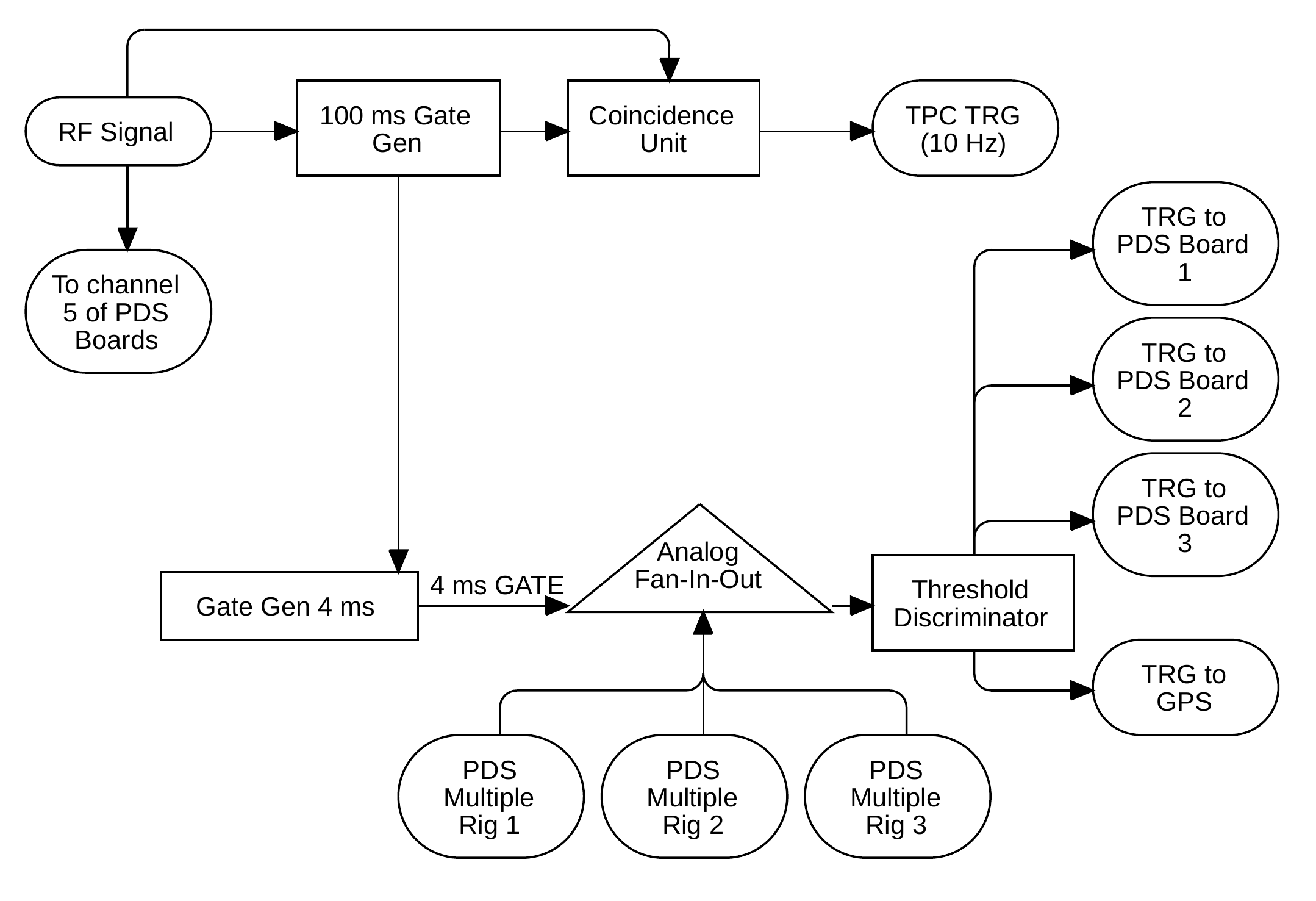}
        \caption{}
    \end{subfigure}%
    \vspace{0.5cm}
    \begin{subfigure}[t]{0.5\textwidth}
        \centering
        \includegraphics[height=2in]{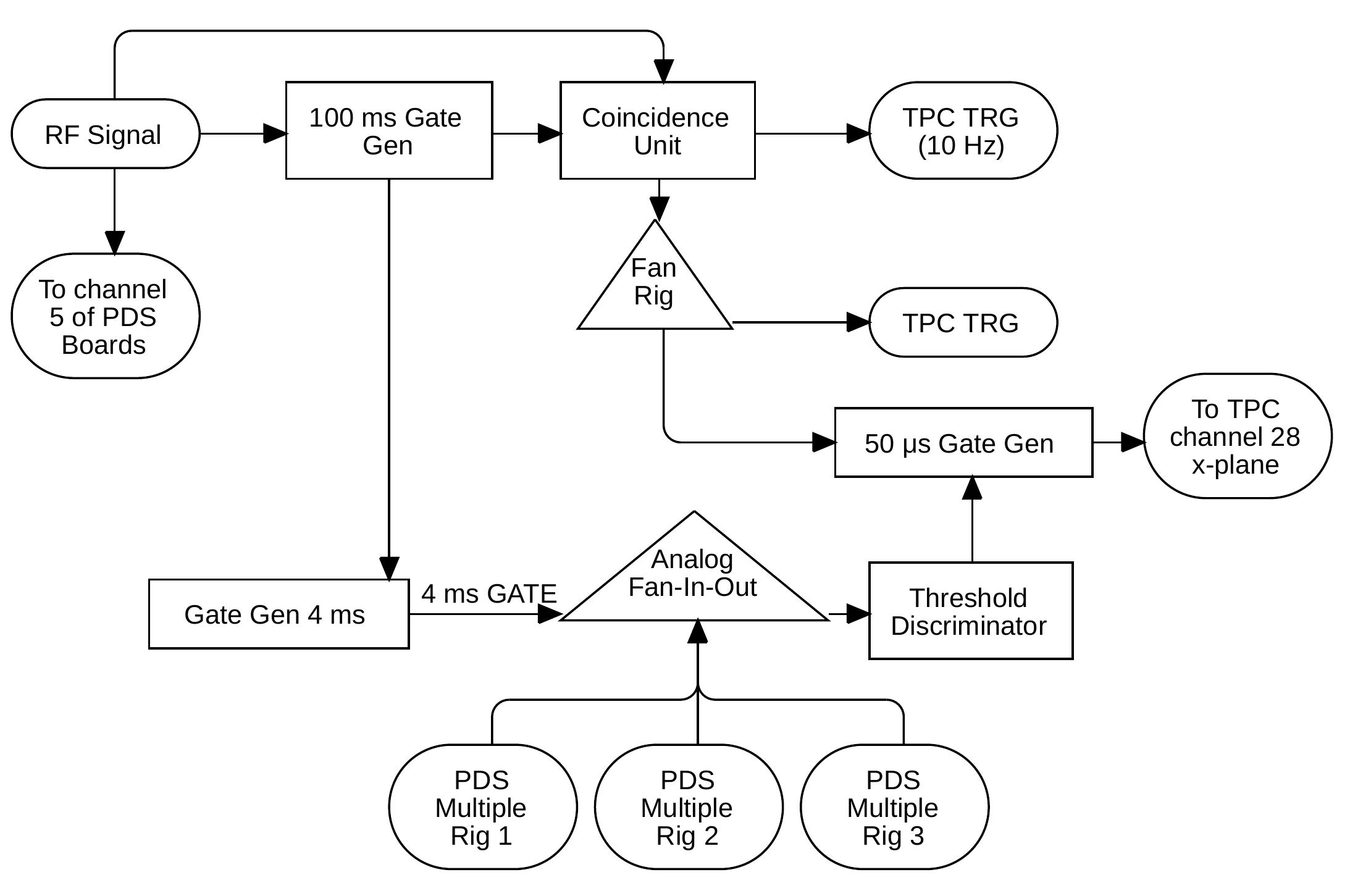}
        \caption{}
    \end{subfigure}
    \caption{Logical flow diagram of the initial trigger design is shown in (a) and the modified trigger design is shown in (b).}
        \label{fig:RF_trigger}
\end{figure}

Flight path 15R is in a building that does not have a clean power source. However, it does have a large power distribution system installed, which provides significant improvement to the noise levels. Isolation transformers were used in a similar manner as was done during the commissioning process described in Section~\ref{sec:miniCAPTAIN_comm}. Both the front-end and back-end electronics racks each had individual isolation transformers, while the clock-module was in its own crate, with an independent power source. 

Following the same principles learned during commissioning, the grounding of the front and back end electronics proved very effective and did not require much modification after the initial liquid argon filling. Some electronic noise did originate from the condenser pump sharing a stand that supports the detector cable tray. Proper grounding of the roughing pump resolved this issue. 

Figure~\ref{fig:grounding_setup} shows a flow chart outlining the final grounding scheme while taking data at the WNR facility at LANSCE. The initial experimental design had the cryostat and front-end electronics isolated from each other. However, after a series of noise studies it was deemed necessary to ground them together. The photon detection system and cryogenic plumbing and feedthroughs were also grounded to the cryostat. The back-end electronics were instead directly grounded to the building ground. It was sufficient to use the grounds of the power source for the computers, calibration fanout, cryogenic main pumps, purity monitors and rack fans. Some extra 2-inch braided straps were used on the neighboring systems (not used by the Mini-CAPTAIN detector), to include a single and three-phase transformer box.

\subsection{Trigger Setup} \label{sec:triggering}

The LANSCE proton beam is delivered in large bunches, called `macropulses', each containing approximately 347 smaller `micropulses'. 
Each macropulse is separated by approximately 8.3\,ms while individual micropulses are separated by 1.8\,$\mu$s. 
The macropulse has a repetition rate of 100\,Hz. The TPC is forced to trigger on the first micropulse in a macropulse, and the TPC DAQ collects data for the duration of  macropulse. After each TPC trigger, subsequent triggers are vetoed for at least half a second, and as long as two seconds, depending on a predetermined DAQ trigger rate.  Figure~\ref{fig:trig_oscope_01} illustrates the timing between the beam RF signal, PDS activity, PDS trigger, and TPC acquisition gate as seen on an oscilloscope.  The frequency of PMT hits demonstrates that at any given time within the TPC DAQ gate at least one PMT is firing.  
However, few neutrons can make it to the detector with the shutter mostly closed, and the resulting neutron interactions produce events with multiple PMT hits.  

The PDS signals are fanned into the analog threshold discriminator. The threshold of each PMT signal was set to 2.5\,mV ($\sim$4 p.e.), and the analog threshold discriminator threshold is set at 800\,mV (corresponding to at least four PMTs firing). This trigger signal is distributed to the GPS and three PDS DAQ boards and is vetoed outside the 4\,ms TPC acquisition window. The first four channels of each PDS DAQ board record the PMT signals, while the fifth channel receives the RF signal from the accelerator.

The PDS trigger is also sent to one of the channels on the TPC DAQ output. The NIM logic pulse from the trigger is stretched into a 50\,$\mu$s signal, which is converted into a differential signal through a NIM/ECL converter and fed to channel 28 on the x-plane. Then the signal is processed through the electronics backplane, where it is transformed into a bipolar pulse with a long tail.  Whenever multiple triggers overlap they produce pileup signals.  Both of these artifacts can be seen in Figure~\ref{fig:PDS_trigger_in_TPC_output}. This input signal is used to help align the PDS and TPS data streams in the offline analysis.

The triggering scheme is shown in Figure~\ref{fig:RF_trigger}. The accelerator RF signal opened a 100\,ms gate generator and triggers the TPC DAQ. Its signal is recorded on channel 5 on each of the PDS digitizers. With the 100 ms gate open, the trigger is sent to a 4 ms gate.  The fan-in-out takes the integrated charge from the individual PMT counters and passes them to the discriminator.  If the summed value exceeds the threshold, the trigger is passed to the PDS boards and GPS for recording.  

It was later determined that it might be challenging to align the TPC and PDS data, due to timing resolution between DAQ machines. Halfway through the data collection, a new triggering scheme was employed where the PDS trigger for record was shared with one of the TPC's digital channels. This signal is passed through a 50 $\mu$s gate generator in coincidence with the trigger to the TPC DAQ.    

\subsection{Neutron Flux Monitoring}
\begin{figure}[t!]
	\centering
        \includegraphics[width=0.48\textwidth]{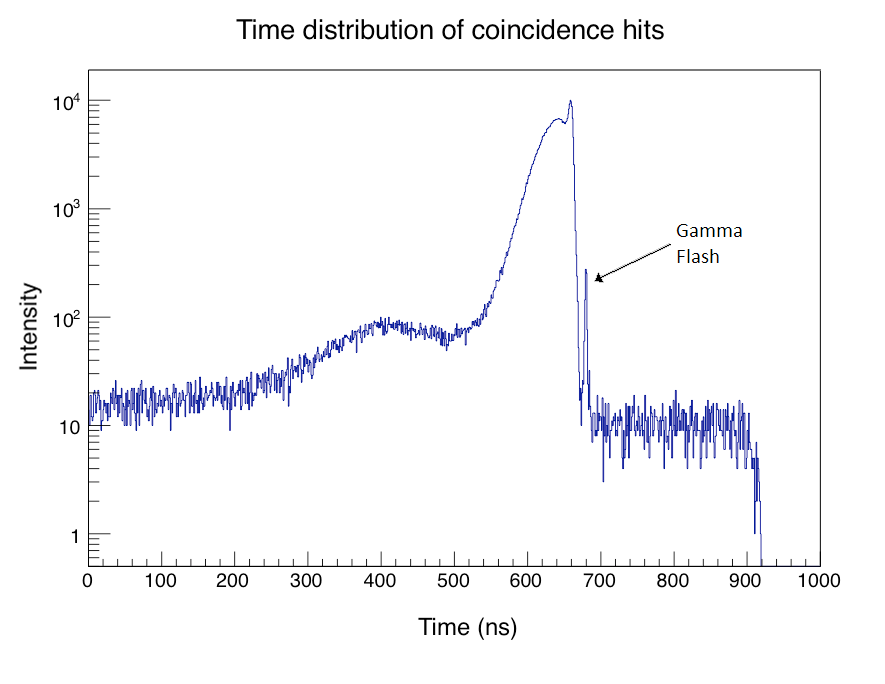}
        \vspace{-0.8cm}
        \caption{Neutron beam flux measured by a plastic scintillator as a function of ($t_\mathrm{beam} - t$).}
        \label{fig:neutron_flux}
\end{figure}

The initial neutron flux was measured with a plastic scintillator detector before positioning Mini-CAPTAIN in the neutron beam.  Measurements were taken with the shutter both completely open and partially shut.  Figure~\ref{fig:neutron_flux} shows the flux with the shutters mostly closed for reduced neutron rate.  The hits in the plastic scintillator detector coinciding with the neutron beam are plotted as a function of ($t_\mathrm{beam} - t$), where $t_\mathrm{beam}$ marks arrival of the neutron beam trigger signal at the digitizer. Note that this parametrization reverses the direction of the time axis. The neutron flux was also measured during the entire Mini-CAPTAIN run cycle. 

\subsection{Mini-CAPTAIN Detector Performance}

\begin{figure*}[h!!]
	\centering
        \includegraphics[trim={3cm 2cm 3cm 1cm}, angle=90,width=0.78\textwidth]{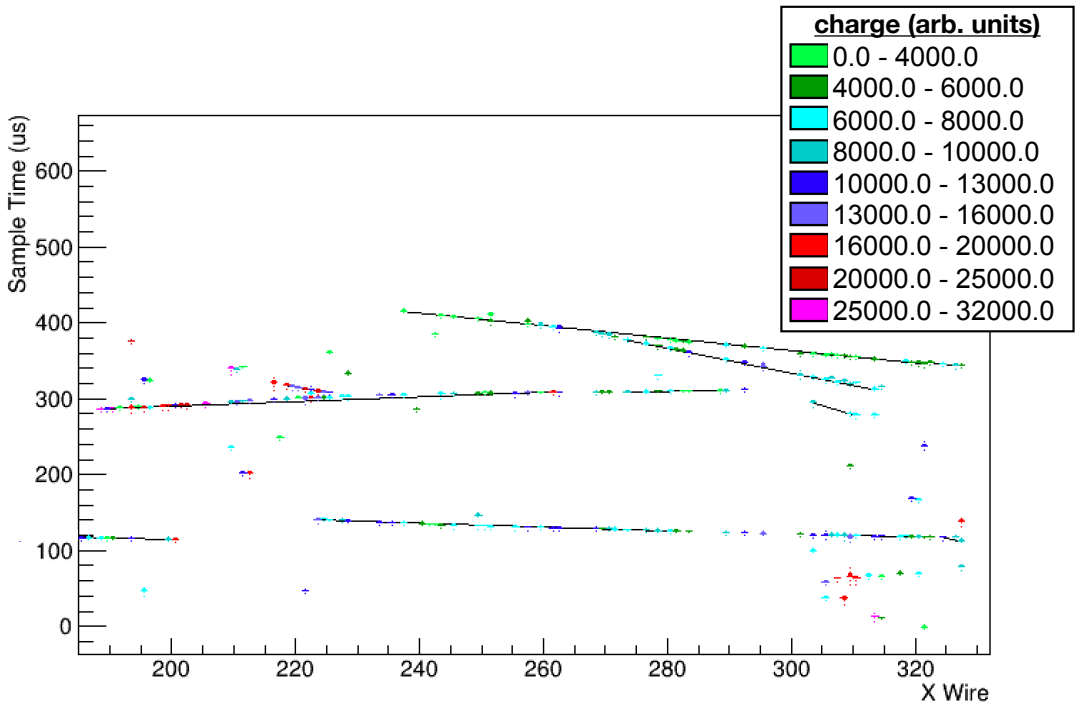}
        \vspace{0.9cm}
        \caption{Neutron interaction candidate tracks observed with the x-plane of the Mini-CAPTAIN TPC in the LANCSE neutron beam at WNR. The color indicates the amount of charge collected on each wire at a given time.}
        \label{fig:x_tracks}
\end{figure*}

The extensive effort to improve the liquid argon purity and minimize noise in the Mini-CAPTAIN detector subsystems allowed for a first measurement of the total neutron cross-section between $100 - 800$\,MeV.  Data-taking in the LANSCE neutron beam with the detector running optimally yields clear evidence of neutron-induced ionization tracks in the TPC.  These candidate neutron events are paired with their corresponding flashes of light observed by the PDS, thereby allowing a calculation of the energy through time-of-flight. Figure~\ref{fig:x_tracks} shows several clear x-plane neutron interaction tracks observed in early data taking at the LANCSE neutron beam. The neutron beam enters the detector from the right side of the plot, the upstream direction corresponding to higher wire numbers. The amount of charge collected on individual wires at a given time is illustrated with an applied color scale as shown in the legend. 

\begin{figure*}[t!] 
    \begin{subfigure}{0.48\textwidth}
    \includegraphics[height=2.1in,width=\linewidth]{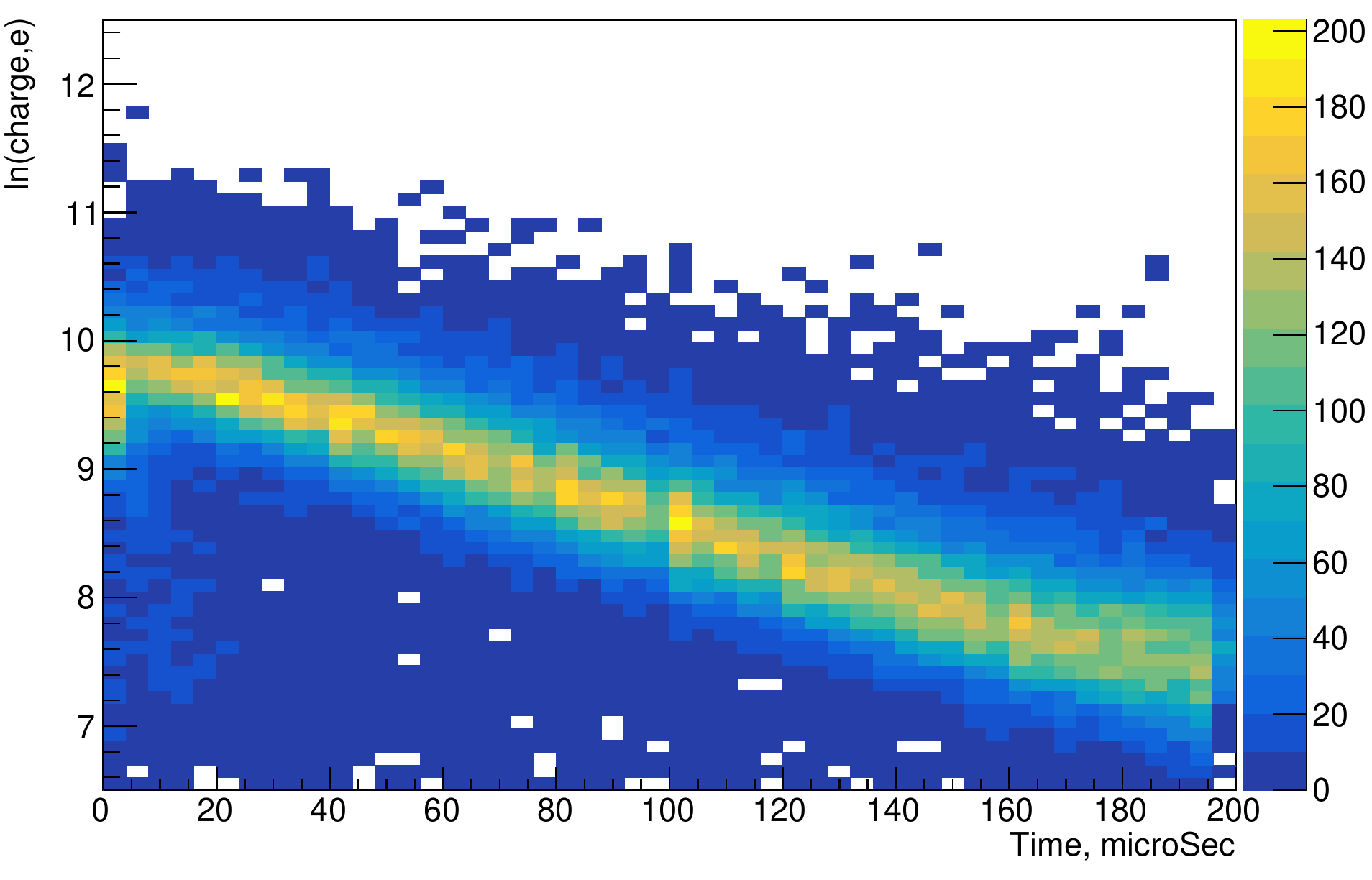}
    \caption{Collected Charge vs Drift Time (x-plane)} \label{fig:a}
    \end{subfigure}\hspace*{\fill}
    \begin{subfigure}{0.48\textwidth}
    \includegraphics[height=2.1in, width=\linewidth]{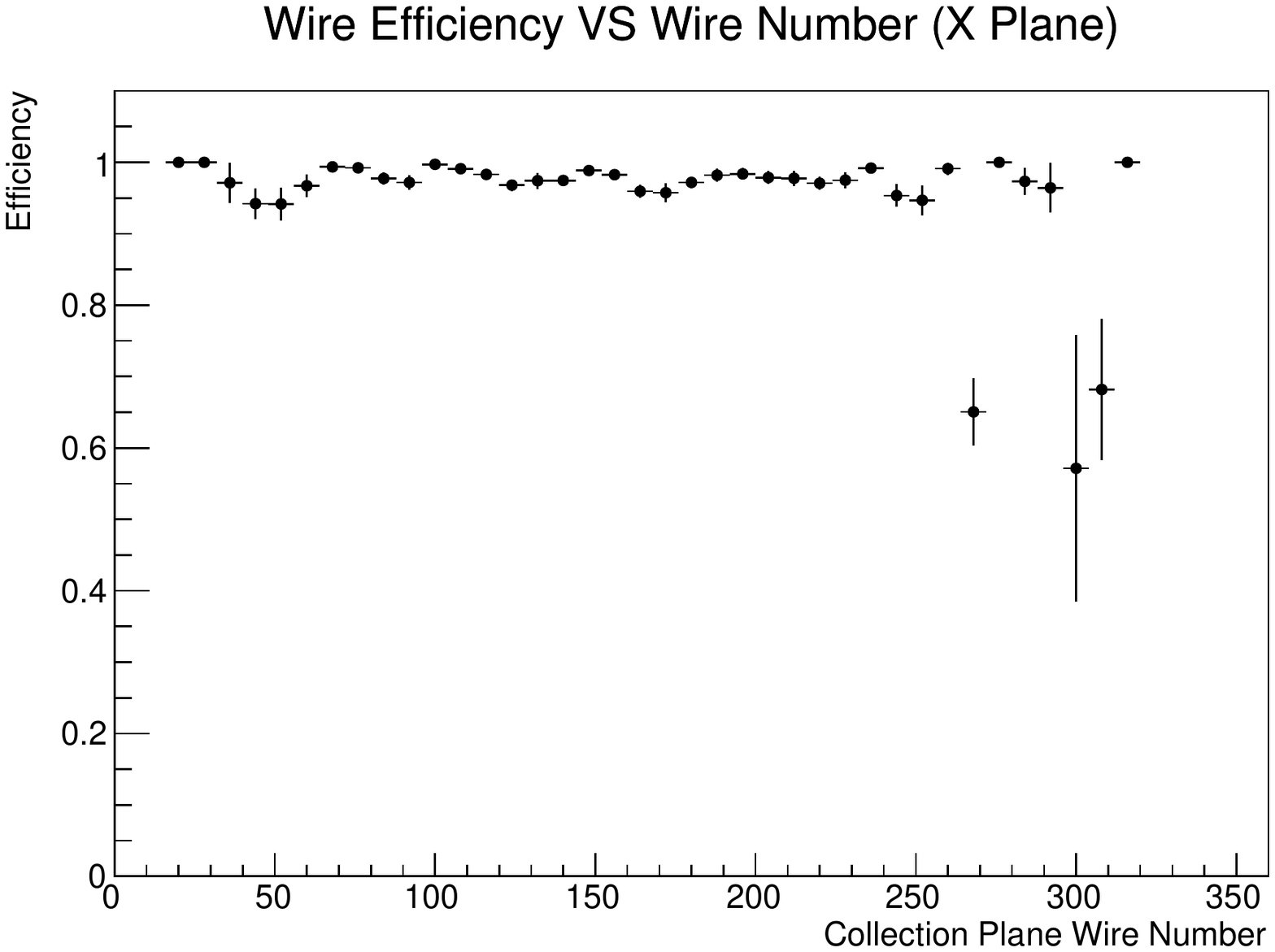}
    \caption{Collection Efficiency vs Wire Number (x-plane)} \label{fig:b}
    \end{subfigure}

    \vspace{1.1cm}
    \begin{subfigure}{0.48\textwidth}
    \includegraphics[height=2.1in, width=\linewidth]{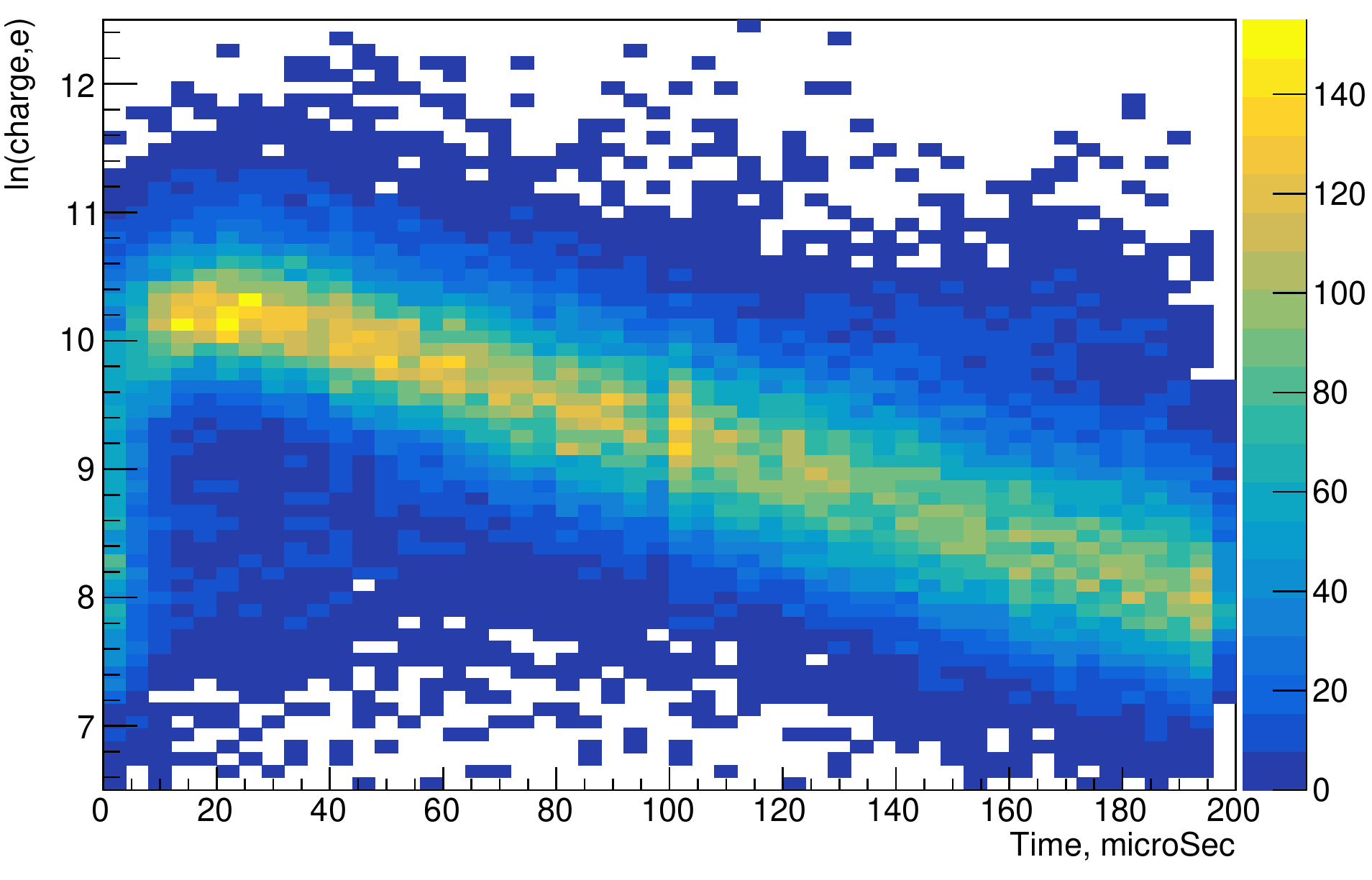}
    \caption{Collected Charge vs Drift Time (u-plane)} \label{fig:c}
    \end{subfigure}\hspace*{\fill}
    \begin{subfigure}{0.48\textwidth}
    \includegraphics[height=2.1in, width=\linewidth]{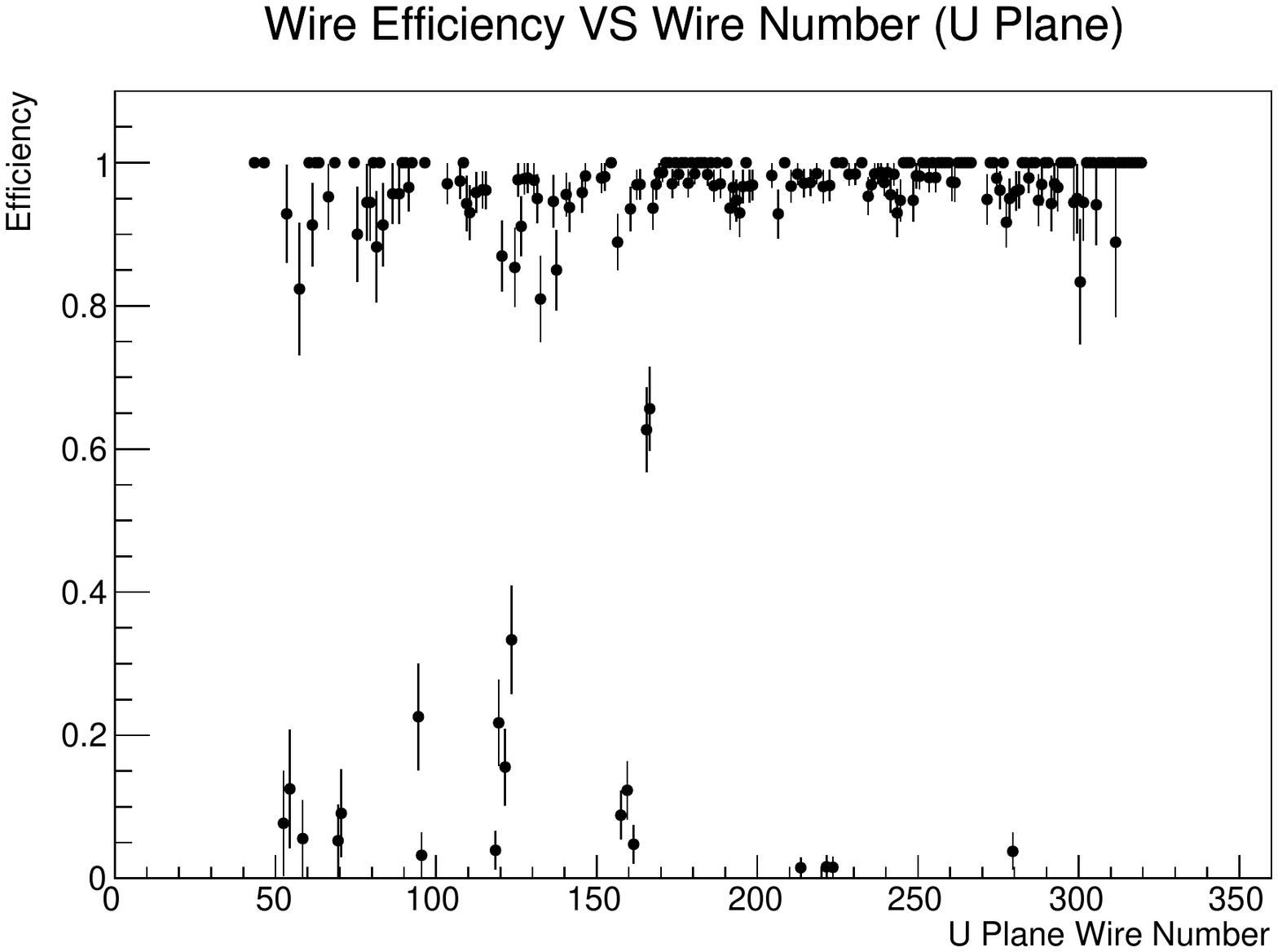}
    \caption{Collection Efficiency vs Wire Number (u-plane)} \label{fig:d}
    \end{subfigure}

    \vspace{1.1cm}
    \begin{subfigure}{0.48\textwidth}
    \includegraphics[height=2.1in, width=\linewidth]{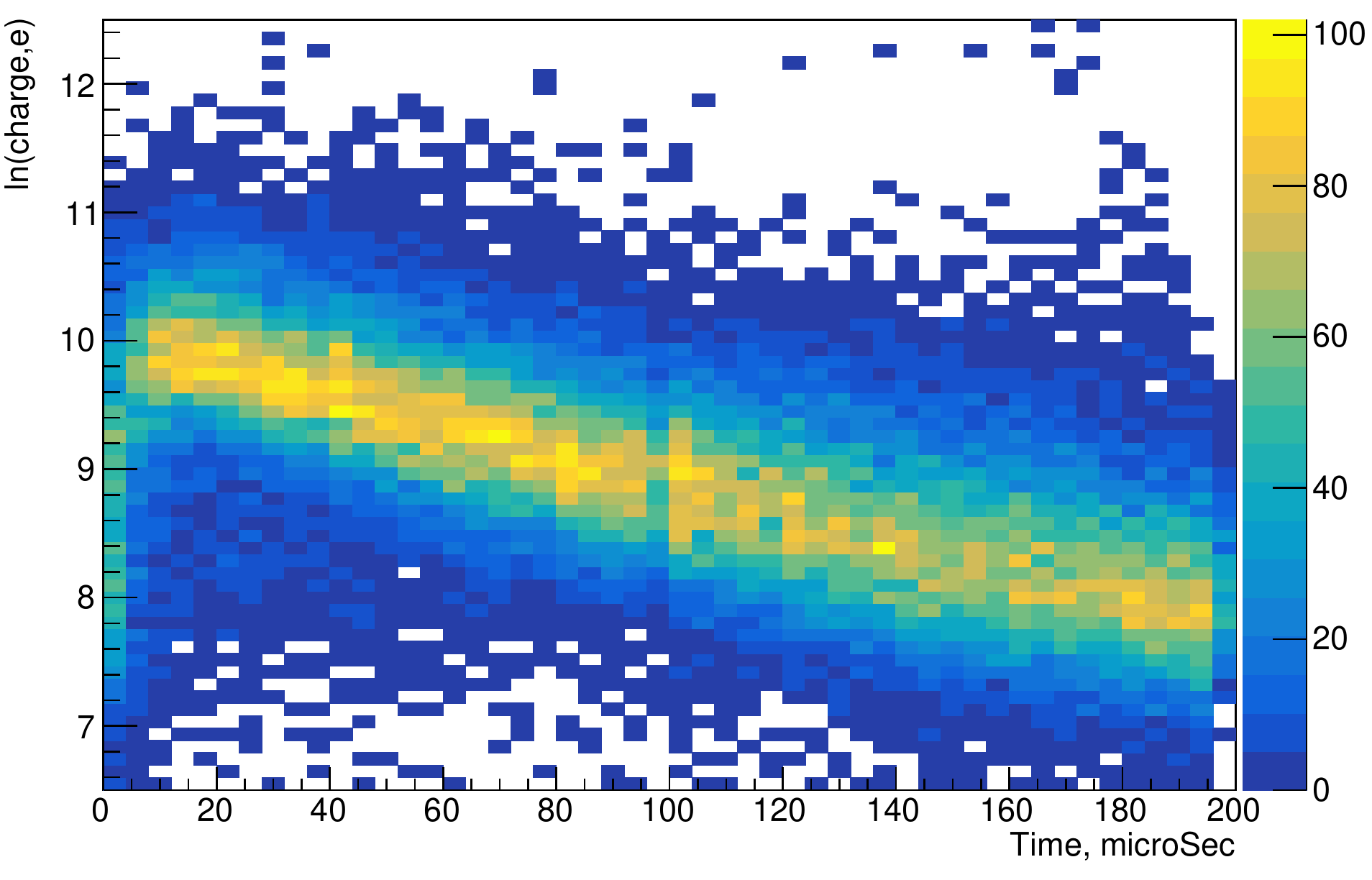}
    \caption{Collected Charge vs Drift Time (v-plane)} \label{fig:e}
    \end{subfigure}\hspace*{\fill}
    \begin{subfigure}{0.48\textwidth}
    \includegraphics[height=2.1in, width=\linewidth]{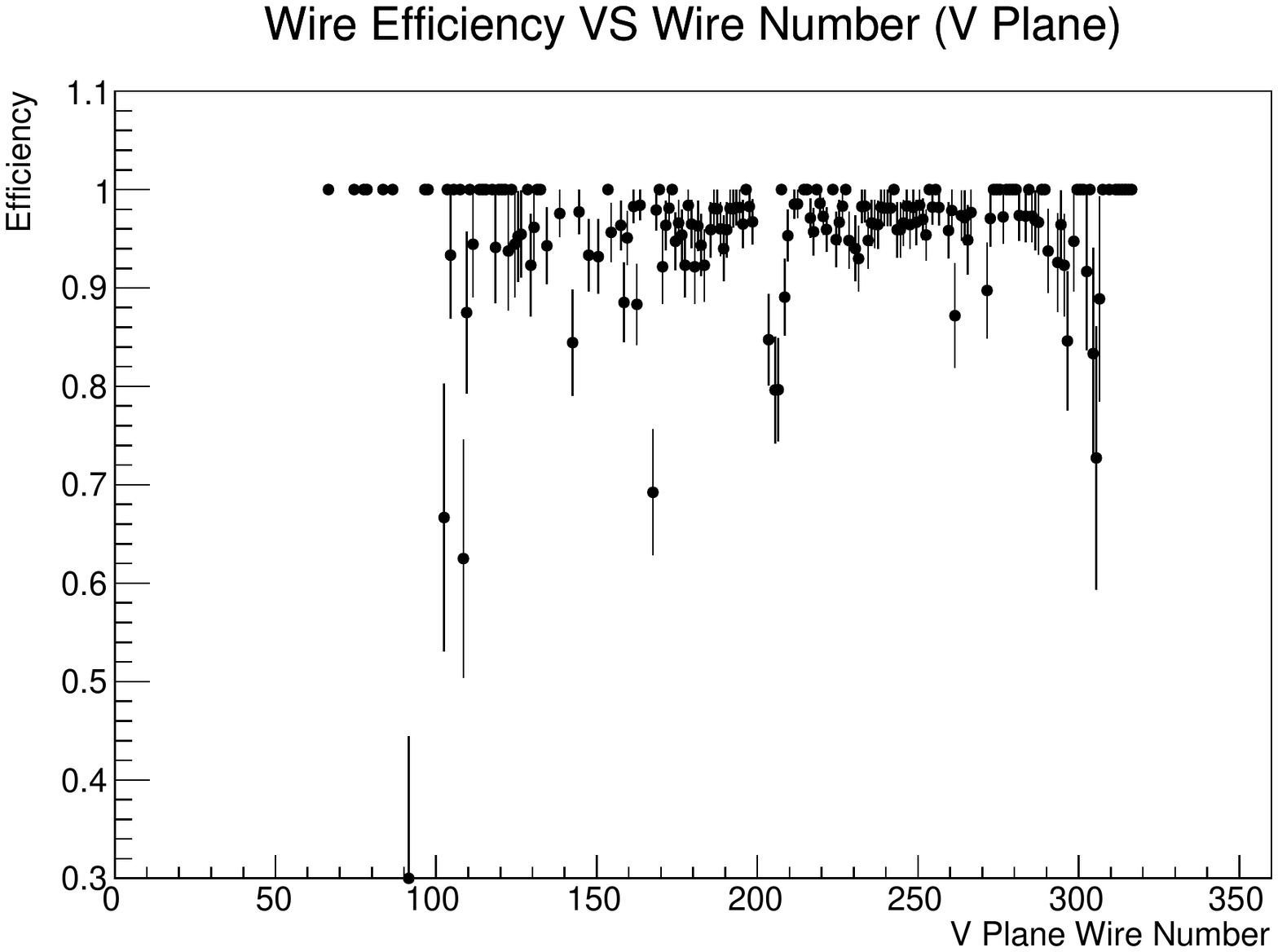}
    \caption{Collection Efficiency vs Wire Number (v-plane)} \label{fig:f}
    \end{subfigure}
    \vspace{0.3cm}

\caption{2D histograms of the charge collected vs the drift time for each wire plain are shown in (a), (c), and (e).  Plots of the calculated wire efficiencies for each wire plane are shown in (b), (d), and (f).} \label{fig:1}
\end{figure*}

Reconstructed cosmic muon tracks spanning the entire TPC drift region, from the cathode plane to the induction planes, are used to estimate the wire collection efficiencies.  This does not require the precision timing information offered by PDS, nor does it rely on the absolute time of a track. Instead the wire plane signals due to drifting electrons are measured as a function of the time between consecutive wire hits. The first wire observing a signal defines the starting time of the track readout period. Since each track is nearly 32 cm long, and the electron drift speed is 1.6 mm/$\mu$s for a 500 V/cm applied field, the last wire hit for a single track arrives 200\,$\mu$s after the first hit.  Additionally, only tracks that crossed at least 30 wires in the x-plane were selected to ensure better track reconstruction.  The wire signals for a track are measured over a 200\,$\mu$s interval divided into 50 time bins, and the collected charge is calculated as the sum of charges of unique wire hits in each time bin. The muon track angle $\theta$ with respect to the wire planes includes a correction factor of $\cos (\theta)$ that is applied to the measured charge on each plane thereby providing absolute collected charge.  The analysis was done separately for the collection wire plane (x) and the induction wire planes (u and v).  The collected charge is plotted as a function of drift time for each plane in the left column of Figure~\ref{fig:1}.  The collection of charge spanning the length of the drift region provided additional evidence that the liquid argon in the TPC was sufficiently purified. 


Within the total drift time interval of 200 $\mu$s, the window between 92--120 $\mu$s corresponds to the location of the neutron beam spot passing through the TPC.  The wire collection efficiency was calculated inside the beam time window by selecting pairs of wire hits, one wire hit occurring immediately before the window and another hit occurring immediately after.  Next, the total number of expected hits inside the time window, $N_{pred}$, is determined, and the predicted hit times of the wires in between the selected pairs is calculated using the electron drift speed. The actual observed number of wire hits inside the beam time window is denoted by $N_{obs}$. The wire collection efficiency is given by $\epsilon = N_{obs}/N_{pred}$.  The collection efficiencies as a function of wire number are shown for each wire plane in the right column of Figure~\ref{fig:1}.

The efficiency study was critical for defining channels that could provide quality data during neutron data taking. The x-plane was observed to have 83 wires in the upstream region of the beam (increasing wire number) that did not meet data quality requirements, compared to 18 wires in the downstream region of the TPC.  Tracks that started in the upstream region of the TPC were ignored when analyzing the first neutron data, thus providing a fairly uniform track reconstruction efficiency for all other events.

\section{Conclusion} \label{sec:conclusion_neut}
The construction, commissioning and successful deployment of the Mini-CAPTAIN detector in a high-energy neutron beam at LANSCE was the result of an extensive campaign lasting several years.  The detector commissioning included several run cycles, each demonstrating that TPC cold electronics, once properly configured, can run stably and efficiently. The electronic modules from both Brookhaven National Lab and Nevis Lab have proven very consistent, with fast and clean signals. Measurement of minimizing ionizing particles produced a nominal signal-to-noise of roughly 9 before any software filter was applied.  The successful extraction of signals from the TPC, in coincidence with time-of-flight signals from the PDS, resulted in a first neutron cross-section measurement in argon. The experience gained during the Mini-CAPTAIN campaign will prove useful in the future operation of CAPTAIN.



\section*{Acknowledgments} \label{sec:Acknowledgments}
This work was supported by the Laboratory Directed Research and Development program of Los Alamos National Laboratory under project numbers 20120101DR and 20150577ER. This work benefited from the use of the Los Alamos Neutron Science Center, funded by the US Department of Energy under
Contract No. DE-AC52-06NA25396 and we would like
to thank Nik Fotiadis, Hye Young Lee and Steve Wender 
for assistance with the 4FP15R beamline.
We gratefully acknowledge the assistance of Mark Makela and 
the P-25 neutron team. D.L.D. acknowledges his support as a Fannie and John Hertz Foundation Fellow.
We further acknowledge the support of the US Department of 
Energy, Office of High Energy Physics and the University of 
Pennsylvania.

\bibliography{CAPTAIN_NIM_Paper}

\begin{thebibliography}{10}
\expandafter\ifx\csname url\endcsname\relax
  \def\url#1{\texttt{#1}}\fi
\expandafter\ifx\csname urlprefix\endcsname\relax\def\urlprefix{URL }\fi
\expandafter\ifx\csname href\endcsname\relax
  \def\href#1#2{#2} \def\path#1{#1}\fi

\bibitem{bib:ICURUS}
E.~Segreto, the ICARUS~Collaboration, Experimental search for the lsnd anomaly
  with the icarus lar-tpc detector in the cngs beam, J Phys.: Conf. Ser. 447
  (2013) 012064.

\bibitem{bib:CaptProposal}
C.~M. et~al., The captain detector and physics program (2013).

\bibitem{bib:neutronMeas}
B.~B. et~al.,
  \href{https://link.aps.org/doi/10.1103/PhysRevLett.123.042502}{First
  measurement of the total neutron cross section on argon between 100 and 800
  mev}, Phys. Rev. Lett. 123 (2019) 042502.
\newblock \href {http://dx.doi.org/10.1103/PhysRevLett.123.042502}
  {\path{doi:10.1103/PhysRevLett.123.042502}}.
\newline\urlprefix\url{https://link.aps.org/doi/10.1103/PhysRevLett.123.042502}

\bibitem{bib:Anderson}
C.~A. et~al., The argoneut detector in the numi low-energy beam line at
  fermilab, Journal of Instrumentation (JINST), JINST 7.

\bibitem{bib:Bakale}
G.~Bakale, U.~Sowada, W.~F. Schmidt, Effect of an electric field on electron
  attachmentto sf$_6$, n$_2$0, and 0$_2$ in liquid argon and xenon, J. Phys
  80~(23) (1976) 2556--2559.

\bibitem{bib:Acciarri}
R.~A. et~al., Effects of nitrogen contamination in liquid argon 2010.

\bibitem{bib:Jones}
B.~J. P.~J. et~al., A measurement of the absorption of liquid argon
  scintillation light by dissolved nitrogen at the part-per-million level
  (2013).
\newblock \href {http://arxiv.org/abs/1306.4605} {\path{arXiv:1306.4605}}.

\bibitem{bib:Criotec}
\href{https://www.criotec.com/en}{Criotec impianti s.p.a.}
\newline\urlprefix\url{https://www.criotec.com/en}

\bibitem{bib:Convery}
M.~C. et~al., A device for quick and reliable measurement of wire tension,
  Princeton/BaBar TNDC (1996) 96--39.

\bibitem{bib:Chen}
H.~C. et~al., Readout electronics for the microboone lar tpc, with cmos front
  end at 89kd, JINST 7.

\bibitem{bib:Cheng}
G.~Cheng, G.~K. at~al., Tpc electronics and readout prototype i. i. test
  results, nevis Laboratory Technical Report (2012).

\bibitem{bib:microBooNE}
{MicroBooNE Collaboration}, The microboone technical design report, DocDB 1821.

\bibitem{bib:Radeka}
V.~Radeka, H.~C. et~al., Cold electronics for "giant" liquid argon time
  projection chambers, Journal of Physics: Conference Series 308 (2011) 012021.

\bibitem{bib:uboone_noise}
R.~A. et~al., Noise characterization and filtering in the microbooneliquid
  argon tpc, JINST 12.

\bibitem{bib:Benson}
C.~Benson, G.~O. Gann, V.~Gehman, Eur. Phys. J. C. 78 (2018) 329.

\bibitem{bib:Gehman}
V.~G. et~al., Nucl. Instr. Methods A 654 (2011) 116.

\bibitem{bib:root}
R.~Brun, F.~Rademakers, {ROOT: An object oriented data analysis framework},
  Nucl. Instrum. Meth. A 389 (1997) 81--86.
\newblock \href {http://dx.doi.org/10.1016/S0168-9002(97)00048-X}
  {\path{doi:10.1016/S0168-9002(97)00048-X}}.

\bibitem{bib:GCheng}
G.~Cheng, G.~K. et~al., Tpc electronics and readout prototype test results,
  Tech. Rep. 1148, Nevis Laboratory, technical Report (2012).

\bibitem{bib:HChen}
R.~A. et~al.,
  \href{https://doi.org/10.1088%2F1748-0221%2F12%2F08%2Fp08003}{Noise
  characterization and filtering in the {MicroBooNE} liquid argon {TPC}},
  Journal of Instrumentation 12~(08) (2017) P08003--P08003.
\newblock \href {http://dx.doi.org/10.1088/1748-0221/12/08/p08003}
  {\path{doi:10.1088/1748-0221/12/08/p08003}}.
\newline\urlprefix\url{https://doi.org/10.1088%2F1748-0221%2F12%2F08%2Fp08003}

\end{thebibliography}

\end{document}